\documentclass[preprint,12pt]{elsarticle}
\usepackage[margin=1in]{geometry}
\usepackage{graphicx} \usepackage{amsmath} \usepackage{amsfonts}
\usepackage{amsmath}
\usepackage{amssymb}
\usepackage{comment}
\usepackage{wrapfig}
\usepackage{floatrow}
\usepackage{breqn}
\usepackage{bm}
\usepackage{multirow}
\usepackage{enumitem}
\usepackage{wasysym}
\usepackage{tikz}

\newcommand{\tikzcircle}[2][red,fill=red]{\tikz[baseline=-0.5ex]\draw[#1,radius=#2](0,0) circle;}

\def\tsr{\eta}

\def\textw{\gamma}
\def\tintw{\zeta}

\def\fpt{\emph{findpts}}

\def\dO{\partial \Omega}

\def\dt{ \Delta t }

\def\scriptO{{{\it O}\kern -.42em {\it `}\kern + .20em}}
\def\RR{{{\rm l}\kern - .15em {\rm R} }}
\def\PP{{{\rm l}\kern - .15em {\rm P} }}
\def\L2{{{\sf L}^2}}
\def\H1{{{\sf H}^1}}
\def\PN2{{\PP_{N}-\PP_{N-2}}}

\def\complex{{{\rm C} \kern - .53em {\rm l} \kern + .38em}}

\def\a1{{ | \lambda_{\min} |}}

\def\l1{{   \lambda_{\min}  }}
\def\bhu{{\hat   {\bf u}}}

\def\bff{{\bf f}}
\def\bffe{{\bf f}_e}

\def\bhu{{\hat   {\bf u}}}
\def\bhn{{\hat   {\bf n}}}

\def\bue{{\underline {\bf e}}}

\def\bu0{{\underline {\bf 0}}}
\def\buu{{\underline {\bf u}}}

\def\br{{\bf r}}

\def\bu{{\bf u}}

\def\bphi{{\bm{\phi}}}

\def\bx{{\bf x}}

\def\bxi{{\bm{\xi}}}

\def\bq{{\bf q}}

\def\bust{\acute{\bu}}

\def\bhn{{\bf \hat n}}
\def\btu{{\bf \tilde u}}

\def\ue{{\underline e}}

\def\u0{{\underline 0}}

\journal{Journal of Computational Physics}

\begin{document}

\begin{frontmatter}



\title{Multirate Timestepping for the Incompressible Navier-Stokes Equations in Overlapping Grids}

\author{Ketan Mittal\corref{cor1}\fnref{label1}}
\author{Som Dutta\fnref{label2}}
\author{Paul Fischer\fnref{label1,label3}}
\fntext[label1]{Mechanical Science \& Engineering, University of Illinois at Urbana-Champaign, 1206 W. Green St., Urbana, IL 61801}
\fntext[label2]{Mechanical \& Aerospace Engineering, Utah State University, 4130 Old Main Hill, Logan, UT 84332}
\fntext[label3]{Computer Science, University of Illinois at Urbana-Champaign, 201 N. Goodwin Ave., Urbana, IL 61801}
\cortext[cor1]{Corresponding author. Present Address: CASC, Lawrence Livermore National Laboratory, 7000 East Avenue, Livermore, CA 94550}

\begin{abstract}
We develop a multirate timestepper for semi-implicit solutions of the unsteady
incompressible Navier-Stokes equations (INSE) based on a recently-developed
multidomain spectral element method (SEM) \cite{mittal2019nonconforming}.
For {\em incompressible}
flows, multirate timestepping (MTS) is particularly challenging because of the tight
coupling implied by the incompressibility constraint, which manifests as an
elliptic subproblem for the pressure at each timestep.  The novelty of our
approach stems from the development of a stable overlapping Schwarz method
applied directly to the Navier-Stokes equations, rather than to the convective,
viscous, and pressure substeps that are at the heart of most INSE solvers.
Our MTS approach is based on a predictor-corrector (PC) strategy that preserves
the temporal convergence of the underlying semi-implicit timestepper. We present
numerical results demonstrating that this approach scales to an arbitrary
number of overlapping grids, accurately models complex turbulent flow phenomenon,
and improves computational
efficiency in comparison to singlerate timestepping-based calculations.

\end{abstract}

\end{frontmatter}

\section{Introduction}\label{sec:intro}

Computational simulations, driven by the improved accessibility of high-performance computing resources, have become ubiquitous for modeling and understanding complex flow phenomena. The accuracy of these computational simulations primarily depends on two key aspects of the method used to solve the partial differential equations (PDE) of interest; (i) the spatial discretization and (ii) the temporal discretization. For spatial discretization, various methods like finite difference (FD), finite element method (FEM), finite volume method (FVM), and the spectral element method (SEM) have become popular for representing the solution of the PDE on a discrete set of nodes/volumes/elements covering the domain $\Omega$ \cite{gresho1978time,rosenfeld1988solution,dfm02}. The spatial accuracy of the discrete solution obtained using these methods depends on the size of the local grid/mesh spacing and the order of accuracy of the spatial discretization. Similarly, the temporal accuracy of the solution depends on the timestep size ($\dt$) and the order ($k$) of the timestepper used for temporal integration. Some popular methods for time-integration include the Runge-Kutta (RK$k$) method, the Adams-Bashforth (AB$k$) method, and the backward differentiation formula (BDF$k$) method. The focus of this work is on the temporal integration of the incompressible Navier-Stokes equations for modeling fluid dynamics and heat transfer in complex domains.

For the unsteady Navier-Stokes equations (NSE), the nonlinear convective term is typically treaded explicitly \cite{tomboulides1997} and the maximum allowable timestep size for stable time-integration is determined by the Courant-Friedrichs-Lewy (CFL) number \cite{lewy1928partiellen}.
For a problem in $d$ space dimensions, a {\em local}
CFL at each gridpoint gridpoint ${\bf x_i}$ is defined as,
\begin{eqnarray} \label{eq:cfllocal}
\mbox{CFL}_i &:=&
\left(\,
\sum_{j=1}^d \left| \frac{\left(c_j\right)_i}{\left(\Delta x_j\right)_i}
\right| \, \right) \, \Delta t,
\end{eqnarray}
where, $\left(c_j\right)_i$ is the $j$th component of
velocity and $\left(\Delta x_j\right)_i$ is an approximate grid spacing in the
${\bf e}_j$ direction, and $\dt$ is the timestep size. Using the local
CFL, a global CFL for the mesh is defined as
\begin{eqnarray} \label{eq:cflmono}
\mbox{CFL} &:=& \max_{x_i} \mbox{CFL}_i.
\end{eqnarray}
To within a scaling factor, the CFL is a robust and easily evaluated
surrogate for $\rho(C) \dt$, where $\rho(C)$ is the spectral radius
of the (assumed skew-symmetric) convection operator.  The CFL limit
associated with explicit time-advancement of the convection operator
is governed by the scale factor, $G$, such that
$\rho(C) \dt  =  G \cdot \mbox{CFL}$ and the chosen timestepper.
For example, for second-order centered differences on a one-dimensional
periodic domain with third-order Adams-Bashforth timestepping, we have
$G=1$ and $CFL < 0.72362$.  For Fourier methods, $G=\pi$, and for
the SEM, $G \approx 1.2$ for $N > 10$ \cite{dfm02}.

\begin{figure}
\begin{center}
$\begin{array}{cc}
\includegraphics[width=46mm,height=30mm]{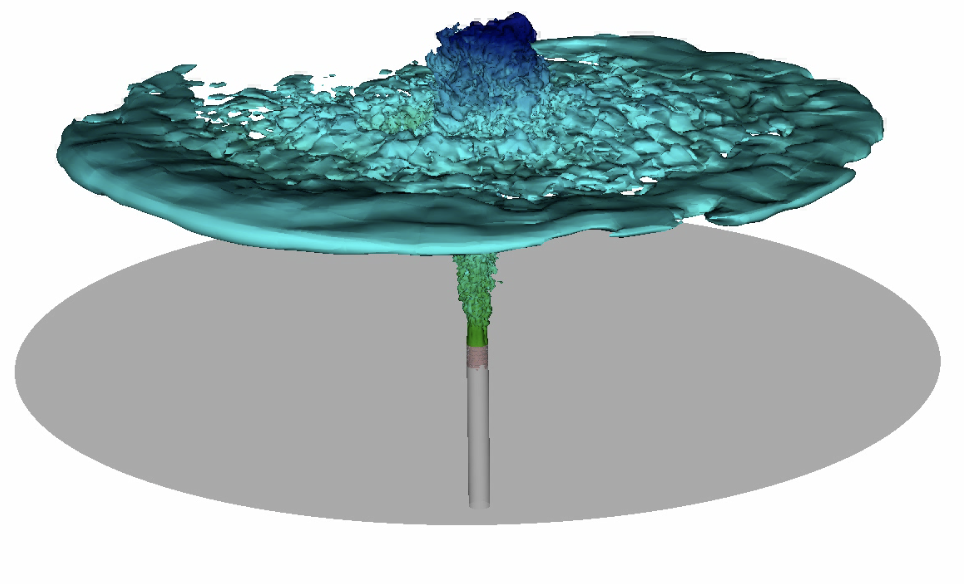} &
\includegraphics[width=45mm,height=45mm]{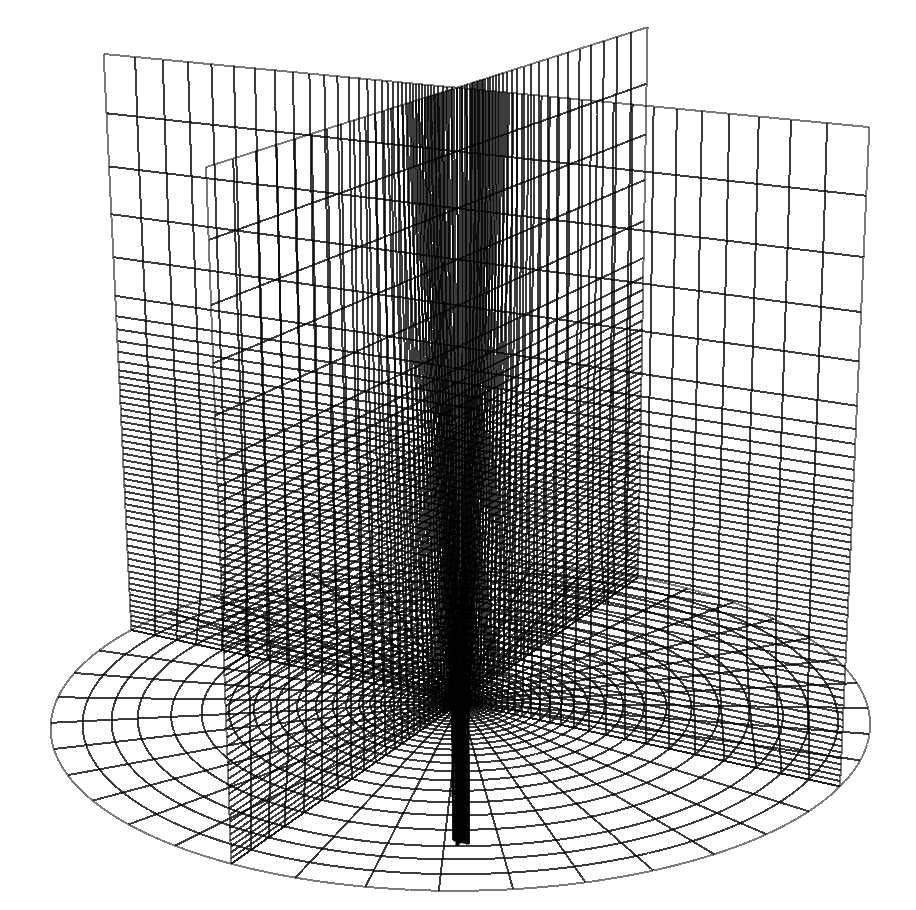} \\
\textrm{(a) Structure of a thermally-buoyant plume} &  \textrm{(b) Monodomain grid} \\
\includegraphics[width=46mm,height=45mm]{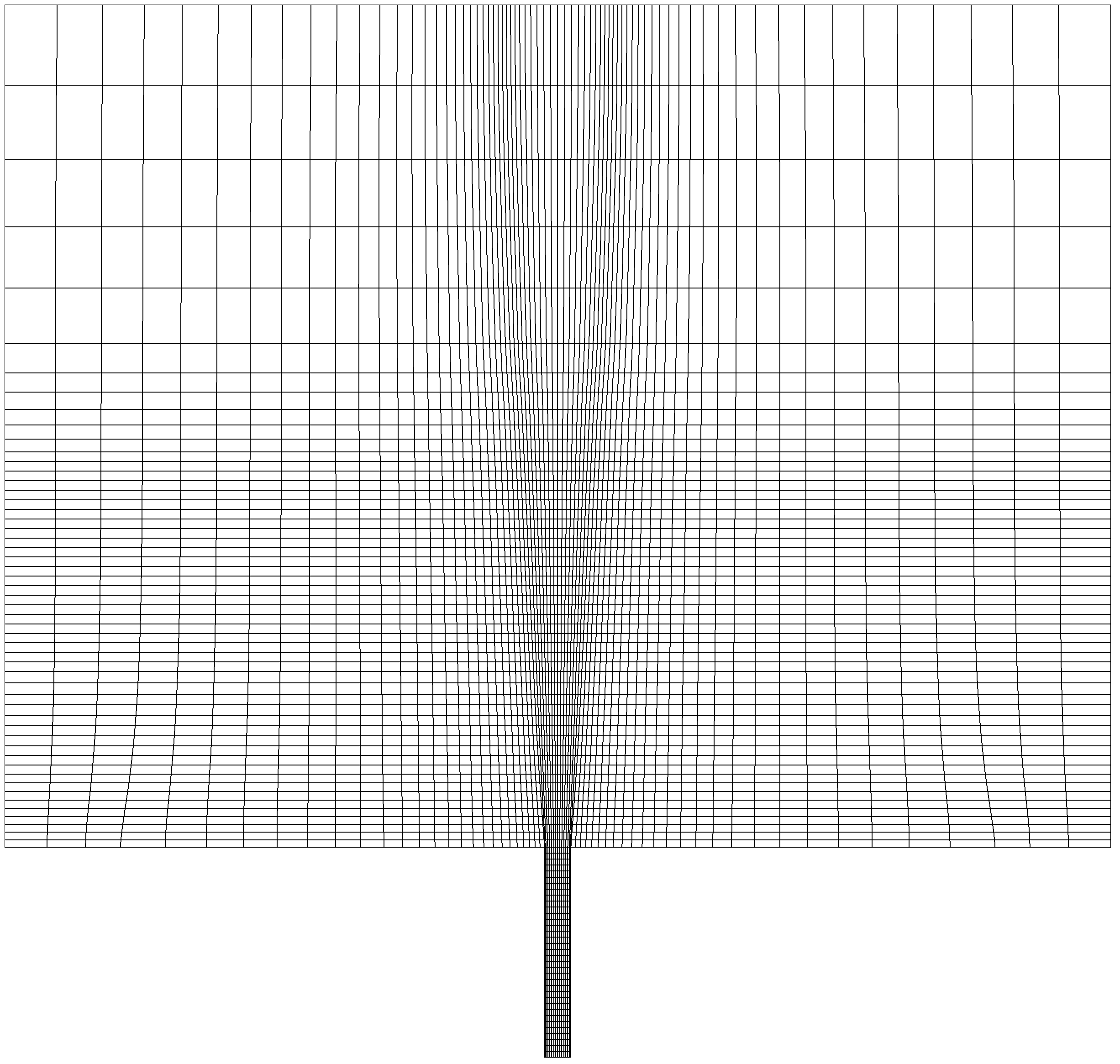} &
\includegraphics[width=45mm,height=45mm]{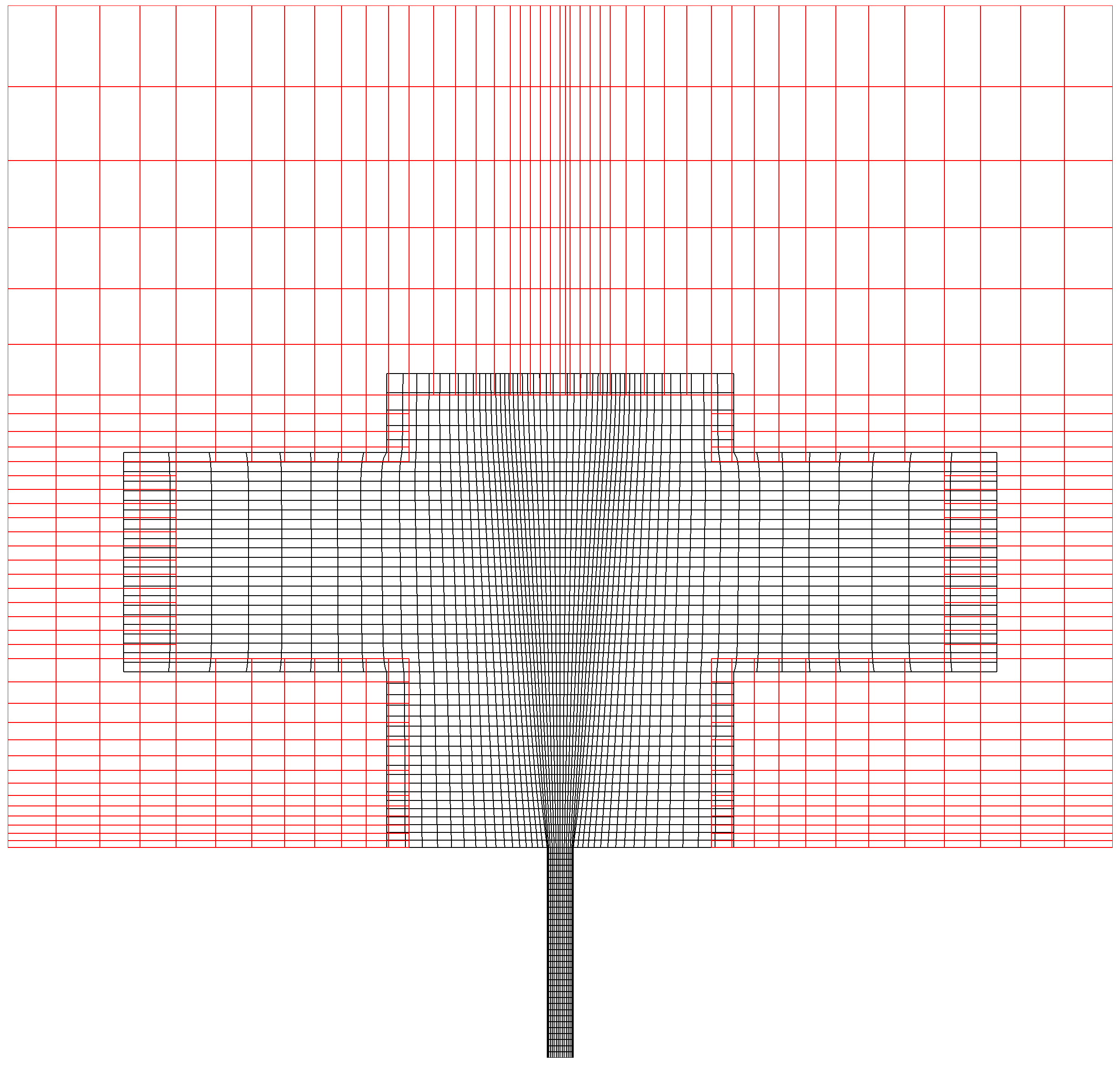} \\
\textrm{(c) Mondomain grid (slice-view)} & \textrm{(d) Overlapping grids (slice-view)} \\
\includegraphics[width=45mm,height=45mm]{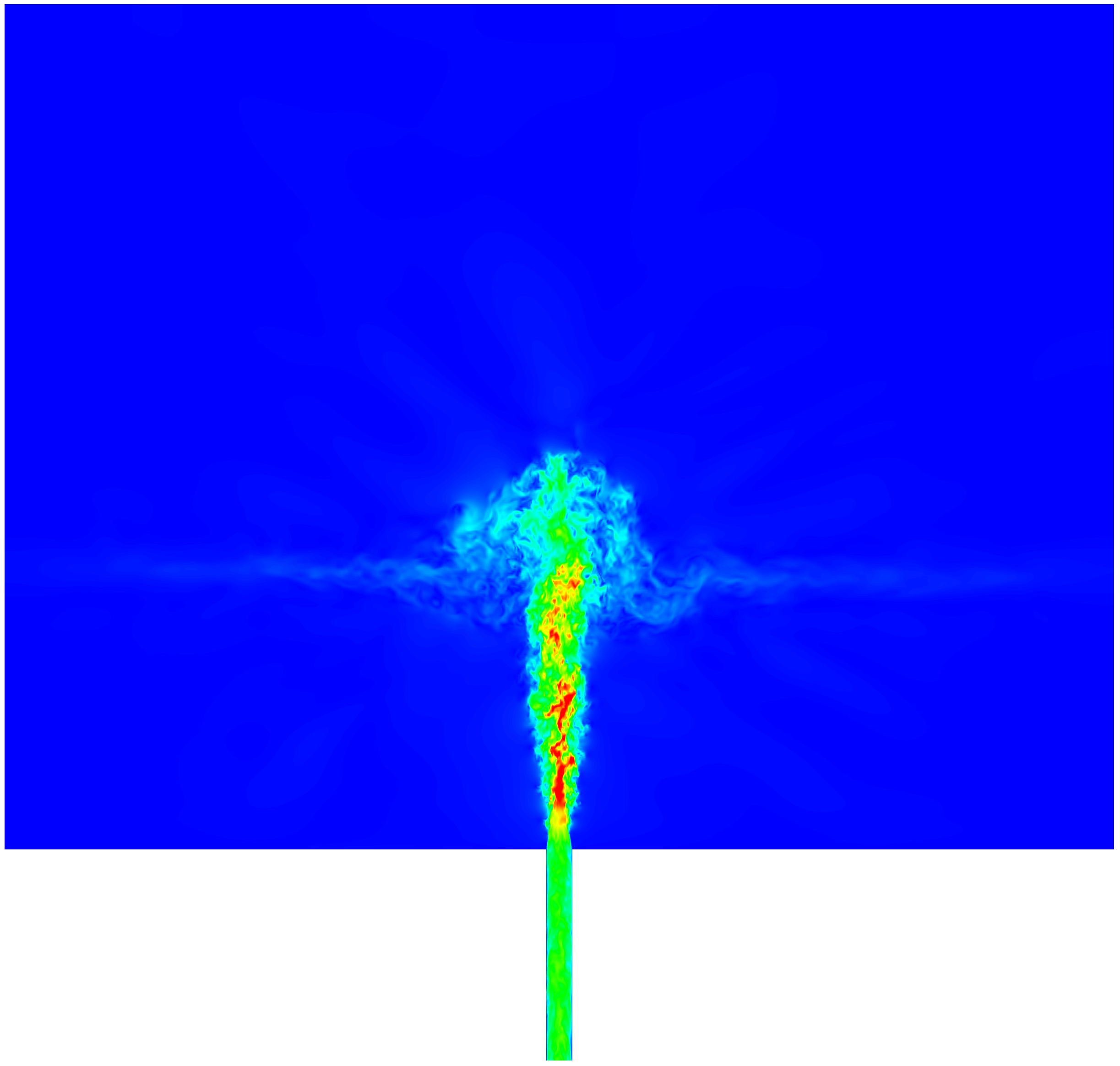} &
\includegraphics[width=45mm,height=45mm]{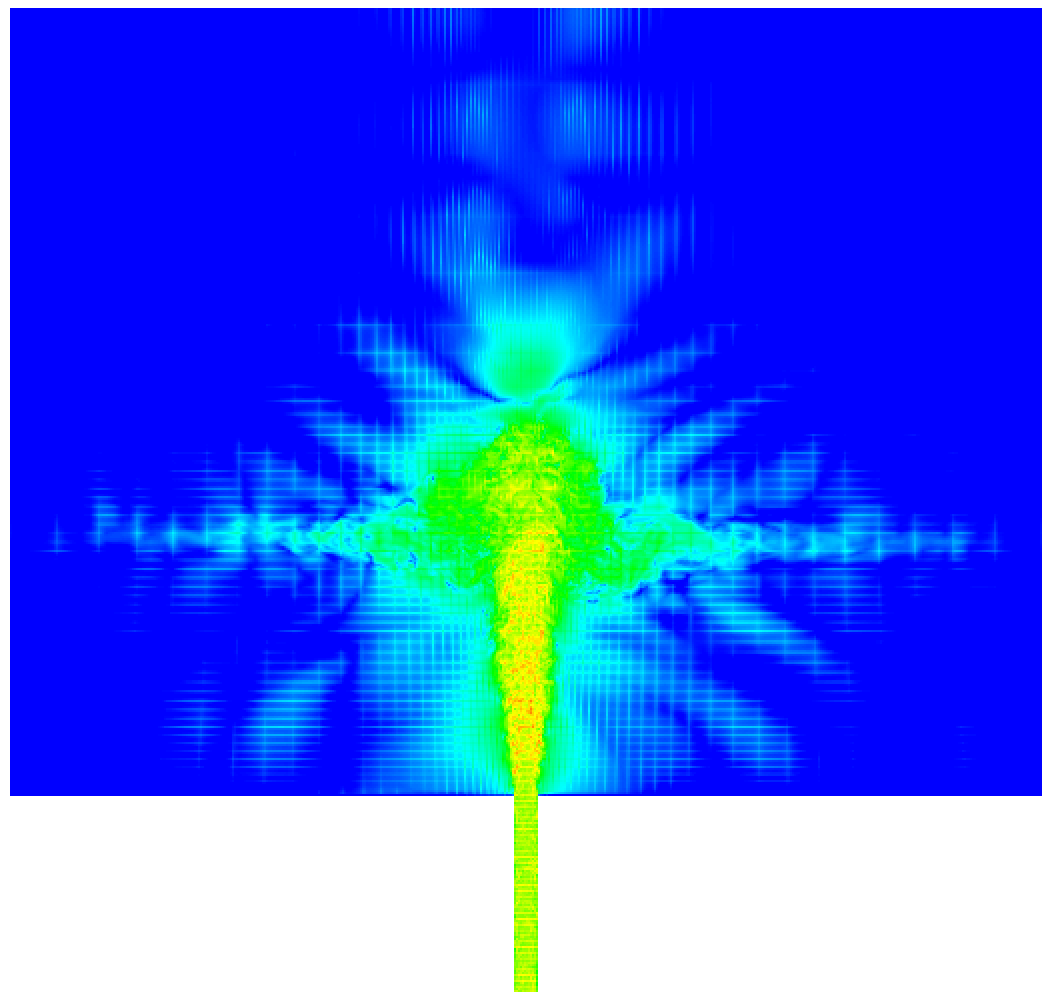} \\
\includegraphics[width=45mm,height=15mm]{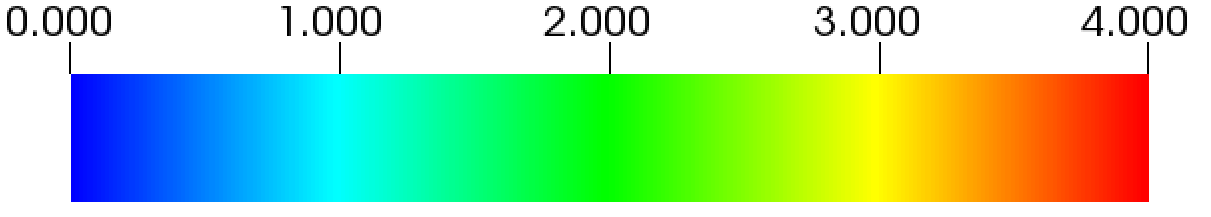} &
 \includegraphics[height=15mm]{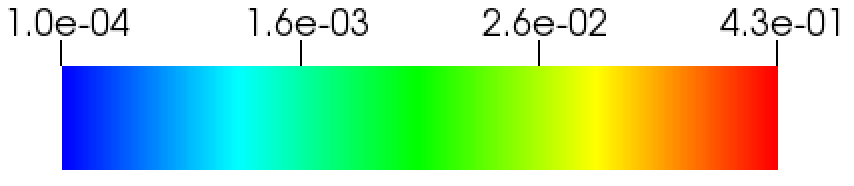} \\
\textrm{(e) Velocity magnitude} & \textrm{(f) CFL} \\
\end{array}$
\end{center}
\vspace{-7mm}
\caption{(a) Structure of a thermally-buoyant plume in a stably stratified environment,
(b) monodomain spectral element mesh,
slice-view of the (c) monodomain mesh, (d) nonconforming overlapping meshes,
(e) velocity magnitude plot, and (f) CFL distribution (log-scale) for a
given timestep size.}
\label{fig:mono_plume}
\end{figure}

Most numerical approximations use the same timestep size throughout the domain.
These methods are classified as singlerate timestepping (STS) methods.
A well-known challenge in STS-based methods, due to the nature of (\ref{eq:cflmono}),
is that even a single point in the domain having a high speed-to-grid-size ratio
can have the undesirable effect of limiting the allowable timestep size throughout
the domain.  This situation occurs, for example, near airfoil trailing edges where
flow speeds are high and computational meshes are often dense.  Another common
case is in the simulation of buoyant plumes \cite{fabregat2016dynamics},
as illustrated in Fig. \ref{fig:mono_plume}.
Figure \ref{fig:mono_plume}(a) shows the structure of a thermally-buoyant plume,
Fig. \ref{fig:mono_plume}(b)-(c) show the monodomain spectral element mesh,
and Fig. \ref{fig:mono_plume}(e) shows the instantaneous velocity magnitude contours. Here,
resolution requirements for the turbulence in and near the inlet pipe result
in having the finest (min $\Delta {\bf x}$) mesh in the region where
flow speed is the highest. Away from the inlet pipe, the turbulence intensity is lower
and the meshes are correspondingly coarser. Consequently, we observe that the
local CFL is almost two to three orders of magnitude higher for the elements
in the plume region as compared to elements in the far-field.
The CFL variation throughout the domain is shown on a log-scale in
Fig. \ref{fig:mono_plume}(f). Using the same timestep size for integrating the NSE,
thus leads to unnecessary computational cost for elements in the far-field.

The scale disparity in CFL is even more evident when nonconforming overlapping grids are used for this domain (Fig.  \ref{fig:mono_plume}(e)). It is well known that overlapping grids are highly effective in reducing the computational cost of calculations for domains featuring flow structures with widely varying spatial scales in different regions of the domain. Since overlapping grids relax the constraint of mesh conformity, they reduce the total element count (10\% in this case) with grids that are constructed according to the physics in the region that they cover.  In this example, since the outer grid is much coarser than the inner grid, the outer grid should be able to use orders of magnitude bigger timestep size than the inner grid. Most overlapping grid-based methods, however, use an STS-based approach \cite{mittal2019nonconforming,henshaw2012cgins,angel2018}, which results in superfluous computational work for the outer grid.

Multirate timestepping methods were first introduced in the seminal work of Rice in 1960 \cite{rice1960split}. Rice developed a Runge-Kutta based timestepping strategy for solving a system of two ordinary differential equations (ODEs) with different integration step sizes. Shortly after, multirate timestepping methods were popularized by Gear \cite{gear1974multirate,gear1984multirate} and Andrus \cite{andrus1979numerical,andrus1993stability}.  In this pioneering work, Gear analyzed the stability and accuracy of Euler method-based strategies for MTS. Similarly, Andrus derived conditions for absolute stability of a high-order Runge-Kutta-based approach for MTS in a system of first-order ODEs. The methods developed by Gear and Andrus were \emph{slowest-first-} or \emph{fastest-first-}based, where the ODE with the slower component is solved first followed by the ODE with the faster component, or vice-versa.
Other similar works include Gunther's multirate partitioned Runge-Kutta-based (slowest-first) scheme for solving a system of ODE with stiff components \cite{gunther2001multirate}, Verhoeven's stability analysis of BDF-based (slowest-first) MTS methods for understanding the time behavior of electrical circuits \cite{verhoeven2007stability}, and Godel's fastest-first Adams-Bashforth-based
scheme for simulation of electromagnetic wave propagation \cite{godel2010gpu}.

While generally useful, a drawback of slowest-first or fastest-first schemes is that they limit the parallelism of the calculation since the ODEs/PDEs are integrated sequentially.  In \cite{engstler1997multirate}, Engstler proposed a method based on Richardson extrapolation to simultaneously solve for the slow and the fast-moving components of a system of ODEs. Similarly, for PDEs, Dawson \cite{dawson2001high}, Constantinescu \cite{constantinescu2007multirate}, and Seny \cite{seny2013multirate} have developed parallel MTS methods for solving hyperbolic or parabolic equations with different timestep size for different element groups in a mesh, based on the local CFL number.

Other notable developments in the area of MTS methods include \cite{savcenco2007multirate,rybak2015multirate,emmett2014high,trahan2012local,gupta2016multi,mikida2018multi}.  In \cite{savcenco2007multirate}, Savcenco introduced a novel approach for MTS where all the equations in a system of ODEs are first integrated using a large global timestep size everywhere in the domain, followed by error indicators to determine the equations that require a smaller timestep size. This approach thus avoids unnecessary computation by using a smaller timestep size only for ODEs that require it. Rybak \cite{rybak2015multirate} has proposed an MTS method for solving fluid flow in coupled free flow domain and porous media. In \cite{rybak2015multirate}, the PDE for the free flow domain (INSE) is first temporally integrated using a CFL-dependent smaller timestep size, followed by a larger timestep size to
solve the PDE for porous media. While Rybak's and Savcenco's approaches are effective for MTS, they are sequential and lack parallelism, similar to the slowest-first- and fastest-first-based methods. In \cite{emmett2014high}, Emmet has used different timestep sizes for solving fluid motion and relatively stiff chemical mechanism to model compressible reacting flow with complex chemistry. This approach can also be extended to conjugate heat transfer problems where the time scale associated with the energy transfer in fluid and solid medium are very different. Trahan \cite{trahan2012local} has developed a fastest-first approach for solving the shallow water equations in monodomain conforming grids, Gupta et al. \cite{gupta2016multi} use multirate timestepping for modeling subsurface methane hydrate reservoirs, and Mikida et al. \cite{mikida2018multi} solve the compressible Navier-Stokes equations using different timestep sizes in overlapping grids with a fastest-first Adams-Bashforth-based scheme.

A survey of the literature shows multirate timestepping methods have mainly been developed for parabolic and hyperbolic problems \cite{gear1984multirate,verhoeven2007stability,engstler1997multirate,constantinescu2007multirate,seny2013multirate,rybak2015multirate,emmett2014high,trahan2012local,mikida2018multi,klockner2010high}.
MTS methods are virtually nonexistent for the incompressible Navier-Stokes equations because the solution is sensitive to the pressure, which satisfies an elliptic Poisson problem at every timestep \cite{dfm02}.
Since the characteristic propagation speed of pressure perturbations is infinite in incompressible flows,
existing approaches for multirate timestepping do not
extend to a single conforming mesh. Overlapping grids however, decouple the
pressure Poisson solve across the different grids modeling a domain, which allows us to develop a
multirate timestepping method. Note that while the MTS method of \cite{rybak2015multirate}
pertains to INSE, it solves the INSE with a fixed timestep size in the entire domain followed by a
different timestep size for the shallow water equations.

In the current work, we develop a parallel multirate timestepping strategy
where the INSE are integrated simultaneously in all the overlapping grids. This method
circumvents the difficulty of the
global divergence-free constraint through a combination of stable high-order
predictor-corrector time integrators and mass-flux corrections for time
advancement of the unsteady Stokes problems. The nonlinear terms continue
to be treated explicitly in time, as in the case of single conforming domain,
but are now advanced without the widely disparate values in CFL throughout the global
domain.  The method scales to an arbitrary number of overlapping grids and
supports arbitrarily high (integer) timestep size ratio. Additionally, the
approach presented in this paper is agnostic to the spatial discretization
(FEM, FVM, SEM, etc.) and can be readily integrated into existing solvers.

The remainder of the paper is organized as follows. Section \ref{sec:prelim} summarizes the monodomain and overlapping grid-based framework for solving the incompressible Navier-Stokes equations using an STS-based approach \cite{mittal2019nonconforming,tomboulides1997}. Section \ref{sec:method} builds upon the STS-based method to describe the MTS-based approach for solving the INSE in overlapping grids. In Section \ref{sec:apps}, we demonstrate that this novel MTS-based approach maintains the temporal accuracy of the underlying BDF$k$/EXT$k$-based timestepper and accurately models complex turbulent flow and heat transfer phenomenon.  Here, we also demonstrate that multirate timestepping reduces the computational cost of a calculation in comparison to the STS-based approach. Finally, in Section \ref{sec:conclusion} we discuss some directions for future work.

\section{Preliminaries }\label{sec:prelim}
This section provides a description of the singlerate timestepping-based
framework for solving the incompressible Navier-Stokes equations in
mono- and multi-domain settings, with the latter based on
overlapping grids.

\subsection{Governing Equations}
We consider the incompressible Navier-Stokes equations in nondimensional
form,
\begin{eqnarray}
\label{eq:nsemomentum}
 \quad \frac{\partial \bu}{\partial t}  + \bu\cdot\nabla{\bu} = - \nabla{p} +
\frac{1}{Re} \nabla^2{\bu} + \bff\, , \\
\label{eq:nsemass}
\nabla\cdot \bu = 0\, ,
\end{eqnarray}
where $\bu(\bx,t)$ and $p(\bx,t)$ represent the unknown velocity and pressure
that are a function of position ($\bx$) and time ($t$), and $\bff(\bx,t)$ is
the prescribed forcing. Here, $Re=LU/\nu$ is the Reynolds number based on the
characteristic length
scale $L$, velocity scale $U$, and kinematic viscosity of the fluid $\nu$.
In addition to the INSE, we also consider the energy equation
\begin{eqnarray}
\label{eq:nseenergy}
\quad \frac{\partial T}{\partial t}  + \bu\cdot\nabla T =
\frac{1}{Pe} \nabla^2{T} + q_T\, ,
\end{eqnarray}
where $T(\bx,t)$  represent the
temperature solution and $q_T(\bx,t)$ is an energy source term. $Pe=1/(Re\cdot Pr)$ is
the Peclet number, which depends on the Reynolds number and the
Prandtl number. The Prandtl number ($Pr=\nu/\alpha$) is the
ratio of the momentum diffusivity ($\nu$) and the thermal
diffusivity ($\alpha$). The solution of
(\ref{eq:nsemomentum})-(\ref{eq:nseenergy}) also depends on the
initial conditions (for time-dependent problems) and boundary
conditions.

\subsection{Spatial Discretization} \label{sec:spatial}
The multirate timestepping method presented in this work
can be readily used with any spatial discretization such as FDM, FEM, and SEM.
All the results presented in this paper are based on the SEM \cite{patera84},
which is a high-order weighted residual method that combines the
geometric flexibility of finite elements ($\Omega$ is decomposed into E smaller elements)
with the rapid convergence of spectral methods. The basis
functions in the SEM are $N$th-order tensor-product Lagrange polynomials
on the Gauss-Lobatto-Legendre (GLL) quadrature points inside each element.
Due to this tensor-product configuration, all operators in SEM can be expressed in
a factored matrix-free form, which leads to fast operator-evaluation ($O(N^{d+1})$)
and low operator-storage (O($N^d$)). The method requires only $C^0$ function
continuity at element interfaces yet yields exponential convergence of the
solution with $N$, resulting in a flexible method with low numerical dispersion.
The reader is referred to \cite{dfm02} for a detailed description of the
SEM. Note that from a timestepping perspective, the global CFL of each
domain is computed by considering the local CFL associated with all the
GLL quadrature points inside each element \eqref{eq:cfllocal}.
Additionally, throughout this paper, we will focus on the temporal discretization
and assume that spatially the grids have been constructed such that they
adequately capture the physics of the flow in the region that they cover.

\subsection{Solution of the INSE in a Monodomain Grid} \label{sec:monoinse}
In our framework, we solve the unsteady INSE in velocity-pressure form using semi-implicit BDF$k$/EXT$k$ timestepping in which the time derivative is approximated by a $k$th-order backward difference formula (BDF$k$), the nonlinear terms (and any other forcing) are treated with a $k$th-order extrapolation (EXT$k$)\footnote{From here on, we will use $k$ to represent the order of accuracy of our temporal discretization, unless otherwise stated.}, and the viscous and pressure terms are treated implicitly.  This approach leads to a linear unsteady Stokes problem to be solved at each timestep, which is split into independent viscous and pressure (Poisson) updates \cite{tomboulides1997}.

Assuming the solution is known at $t^{n-1}$ and that a constant timestep size $\dt$ is used for all timesteps,
we compute a tentative velocity field at time $t^n$ with contributions from the BDF$k$ and the explicit terms as
\begin{eqnarray} \label{eq:vstarsol}
\bust^n &=& -\sum_{j=1}^{k} \beta_j
\bu^{n-j} + \dt\,\sum_{j=1}^{k} \alpha_{j}
\bffe^{n-j},
\end{eqnarray}
where we use $\bffe$ to represent the explicit contributions
\begin{eqnarray} \label{eq:fe}
\bffe^{n-j} = (-\bu\cdot\nabla\bu+\bff)^{n-j},
\end{eqnarray}
the superscript $(\,\,)^{n-j}$ indicates quantities evaluated at earlier timesteps, $t^{n-j}$, and $\beta_j$ and $\alpha_j$ are the BDF and EXT coefficients, respectively. $\bust^n$ constitutes the nonlinear update but does not account for the divergence-free constraint or viscous effects. The divergence-free constraint (\ref{eq:nsemass}) is enforced through a pressure correction. A pressure Poisson equation is obtained by taking the divergence of the momentum equation, assuming the solution is divergence-free at time $t^n$, $\nabla \cdot \bu^n = 0$, and using the identity $\nabla^2 \bu^n = \nabla(\nabla \cdot \bu^n) - \nabla \times \nabla \times \bu^n$:

\begin{eqnarray} \label{eq:prsol}
-\nabla^2 p^n &=& -\frac{\nabla \cdot \bust^n}{\dt} + \frac{1}{Re} \nabla \cdot
\sum_{j=1}^{k} \alpha_{j} (\nabla \times \boldsymbol\omega^{n-j}),  \\
\implies -\nabla^2 p^n &=& \nabla \cdot \bff_p,
\end{eqnarray}
where $\boldsymbol\omega^n = \nabla \times \bu^n$, and
\begin{eqnarray}
\bff_p &=& -\frac{\bust^n}{\dt} + \frac{1}{Re}
\sum_{j=1}^{k} \alpha_{j} (\nabla \times \boldsymbol\omega^{n-j}).
\end{eqnarray}
The advantage of using the curl-curl form for the viscous term to decouple the velocity and pressure solve is that the equation governing the error in divergence ($\nabla \cdot \bu^n$) is an elliptic PDE instead of a parabolic PDE.  As a result, this formulation is stable with splitting-induced divergence errors that are only $\scriptO(\dt^k)$ \cite{tomboulides1997,fischer17}.

Substituting the pressure solution $p^n$ in (\ref{eq:nsemomentum}), $\bu^n$ is obtained by solving the Helmholtz equation
\begin{eqnarray} \label{eq:velsol}
\frac{\beta_0}{\Delta t}\bu^n -\frac{1}{Re}\nabla^2{\bu}^n = -\nabla{p}^n + \frac{\bust^n}{\dt}.
\end{eqnarray}
Similar to (\ref{eq:velsol}), using implicit treatment of the diffusion term and explicit treatment of the advection term for the energy equation, the solution $T^n$ for temperature is obtained by solving the Helmholtz equation
\begin{eqnarray} \label{eq:tempsol}
\frac{\beta_0}{\Delta t}T^n -\frac{1}{Pe}\nabla^2{T}^n = \sum_{j=1}^{k}
\frac{\beta_j}{\Delta t}\bu^{n-j} + \sum_{j=1}^{k} \alpha_{j} (-\bu\cdot \nabla
T+q_T)^{n-j}.
\end{eqnarray}

Spatial discretization of (\ref{eq:prsol})-(\ref{eq:tempsol}) is
based on variational projection operators \cite{dfm02}. We impose
either essential (Dirichlet) boundary conditions or natural
(Neumann) boundary conditions on a surface for velocity (and
temperature). As expected, surfaces that have Dirichlet
conditions for velocity have Neumann conditions for pressure, and
vice-versa. Note that we use $\dO_D$ to denote the subset of
domain boundary $\dO$ on which Dirichlet conditions are imposed
on velocity and $\dO_N$ for the subset (e.g., outflow) on which
pressure is prescribed.

The Navier-Stokes solution time-advancement can be summarized as:

\begin{enumerate}[leftmargin=*]
\item Compute the tentative velocity field $\bust^n$ using (\ref{eq:vstarsol}), which accounts for the BDF$k$ and time extrapolated nonlinear terms (EXT$k$ terms).

\item Solve the {\em linear} Stokes subproblems (\ref{eq:prsol}) and
(\ref{eq:velsol}) to compute the velocity-pressure solution, $\bphi^n=[\bu^n,p^n]^T$
\begin{eqnarray}
\label{eq:unsteadystokes}
{\bf S} \bphi^n &=& \br^n, \quad \bu^n \rvert_{\dO_D} = \bu_b^n, \quad p^n
\rvert_{\dO_N} = 0.
\end{eqnarray}

\end{enumerate}
Here $\br^n$, determined using $\bust^n$, accounts for all
inhomogeneities for both pressure and velocity, given on the
right-hand sides of (\ref{eq:prsol}) and (\ref{eq:velsol}),
respectively: \begin{equation}
\begin{aligned}
\label{eq:unsteadystokesbr}
\br^n&=[\br_v^n,r_p^n]^T, \\
\br_v^n &= -\nabla{p}^n + \frac{\bust^n}{\dt},\,\,\,\, r_p^n &= -\frac{\nabla \cdot \bust^n}{\dt} + \frac{1}{Re} \nabla \cdot
\sum_{j=1}^{k} \alpha_{j} (\nabla \times \boldsymbol\omega^{n-j}).
\end{aligned}
\end{equation}
In (\ref{eq:unsteadystokes}), $\bu_b^n$ is the prescribed
velocity on all Dirichlet surfaces ($\dO_D$) of the domain,
homogeneous Dirichlet conditions are imposed for pressure on
outflow surfaces ($\dO_N$), and homogeneous Neumann conditions
are imposed for velocity on $\dO_N$. It is straightforward to
show that the Neumann conditions for pressure on $\dO_D$ can be
represented as a function of the Dirichlet condition for velocity
($\bu_b^n$) \cite{tomboulides1997}.

From an implementation perspective, the solution to the unsteady Stokes problem
is obtained in our framework by first solving the pressure Poisson equation \eqref{eq:prsol}
using a three-level $p$-multigrid accelerated by GMRES \cite{lottes2005hybrid},
followed by a diagonally preconditioned conjugate gradient iteration \cite{dfm02}
for the Helmholtz solve associated with each component of the velocity \eqref{eq:velsol}.
Note that for each of these iterative solves, we use a matrix-free approach
enabled by the tensor-product structure of
the basis functions in the SEM. Further details on the matrix-free operator
evaluation with SEM are provided in Section 2.1 of \cite{mittal2019phd}.

In \eqref{eq:unsteadystokes},
we have omitted the solution to temperature (\ref{eq:tempsol})
for brevity since it is similar to the Helmholtz solve for
velocity. Note that since we always impose Dirichlet conditions for
velocity and pressure on $\dO_D$ and $\dO_N$, respectively, we will omit
them in the description of timestepping for overlapping grids.

\subsection{Solution of the INSE on Overlapping Grids} \label{sec:methodschwarz}
\begin{figure}[t!] \begin{center}
$\begin{array}{cc}
\includegraphics[height=43mm]{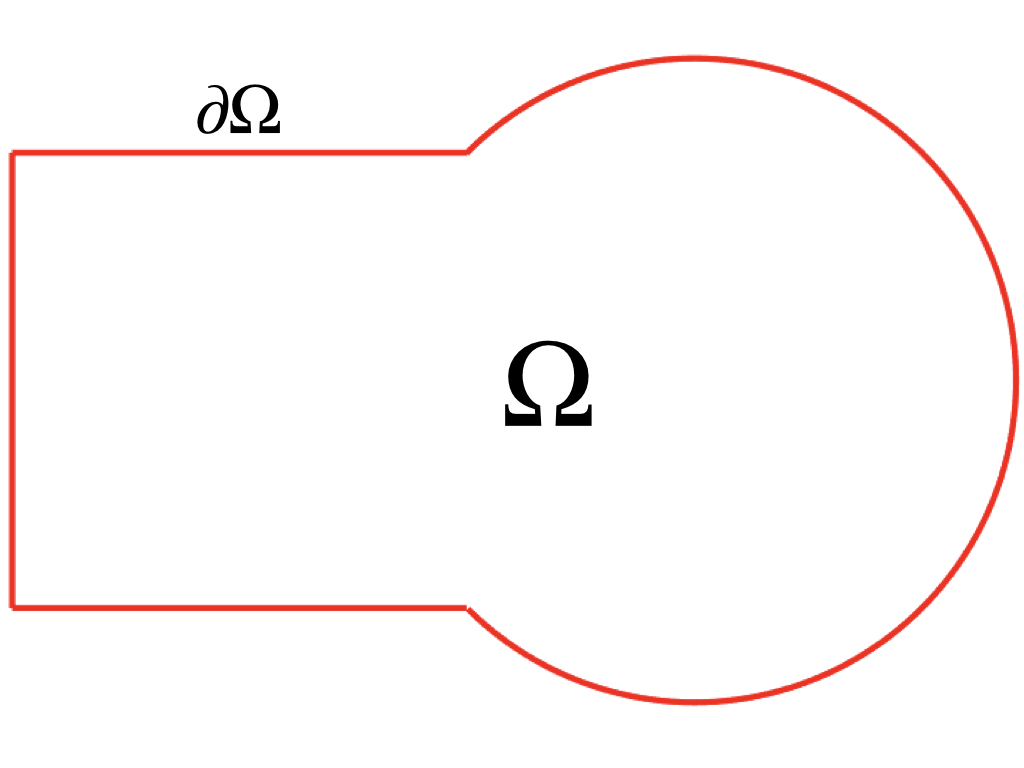} & \hspace{2mm}
\includegraphics[height=45mm]{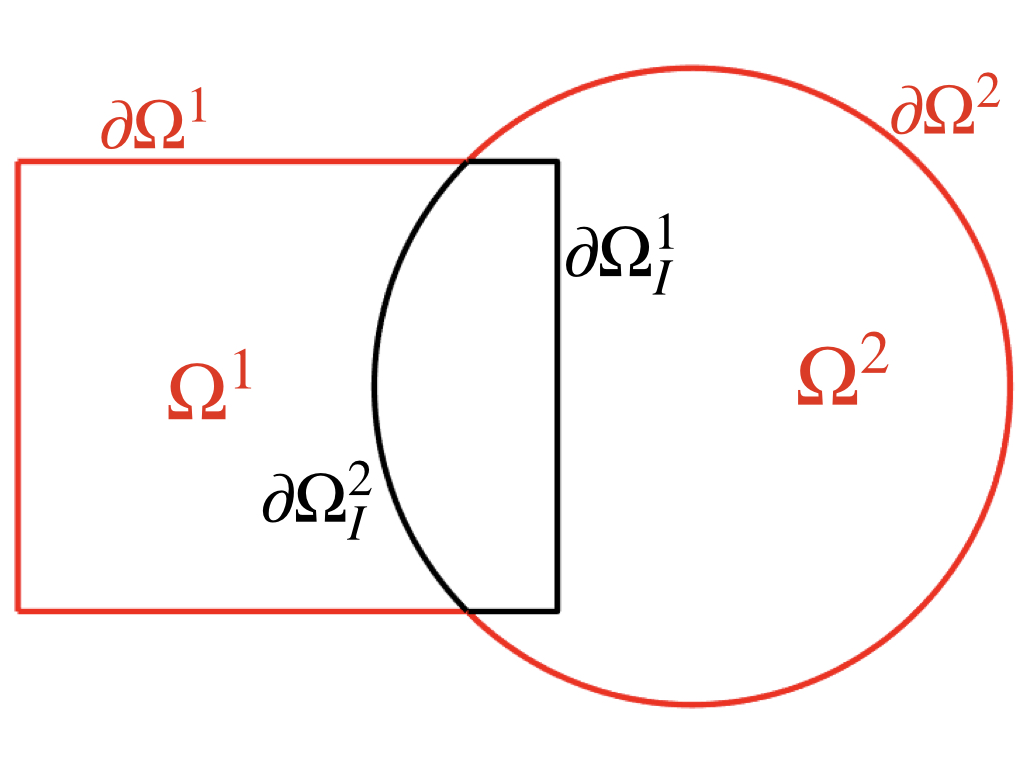}
\end{array}$
\end{center}
\vspace{-7mm}
\caption{(left to right) (a) Composite domain $\Omega$ (b) modeled by overlapping
rectangular ($\Omega^1$) and circular ($\Omega^2$) subdomains. $\dO^{s}_{I}$
denotes the segment of the subdomain boundary $\dO^s$ that is interior to
another subdomain $\Omega^r.$}
\label{fig:schwarz}
\end{figure}

The overlapping Schwarz method for solving a PDE in overlapping domains was introduced by Schwarz in 1870 \cite{schwarz1870}. The decomposition for Schwarz's initial model problem is illustrated in Fig. \ref{fig:schwarz} where the domain $\Omega$ is partitioned into two subdomains, a rectangle ($\Omega^1$) and a circle ($\Omega^2$), with nonzero overlap such that $\dO^1_{I}:=\dO^{1} \subset \Omega^2$ and $\dO^2_{I}:=\dO^{2} \subset \Omega^1$. We use $\dO^s_I$ to denote the ``interdomain boundary'', namely the segment of the subdomain boundary $\dO^s$ that is interior to another subdomain. The interdomain boundaries $\dO^1_{I}$ and $\dO^2_I$ are highlighted in Fig. \ref{fig:schwarz}(b).

There are two key aspects for solving a PDE (INSE) in
overlapping grids. First, since overlapping grids introduce
interdomain boundaries, a robust mechanism is required to interpolate
boundary data for the gridpoints discretizing $\dO^s_I$ from
the subdomain $\Omega^r$ that they overlap. Second, Schwarz iterations
are required to ensure that the solution is consistent across the different
overlapping grids.

There are two popular approaches for Schwarz iterations. In the alternating
Schwarz method, given $S$ overlapping
subdomains, the PDE in solved in the first subdomain and that solution is used to
update the interdomain boundary data in all other subdomains.
This process is repeated sequentially for $s=2\dots S$ subdomains.
A drawback of the alternating Schwarz method is that it does not scale with the
number of subdomains since it requires at-least $S$ steps to obtain the solution of a
PDE. In contrast, the simultaneous Schwarz method solves
the PDE simultaneously in all subdomains followed
by interdomain boundary data exchange. This iterative process is repeated
until the solution converges to desired accuracy in the overlap region. Thus,
assuming there is a robust mechanism to effect interdomain
boundary data exchange, the scalability of the simultaneous Schwarz iterations is not
restricted by the number of subdomains. The Schwarz-SEM framework that we describe
next is based on the simultaneous Schwarz method, and the reader is
referred to \cite{smith2004domain} for additional details on different OS-based
techniques.

For notational purposes, we introduce $\bphi^{s,n,q}$ as the
solution on the $q$th Schwarz iteration in subdomain $\Omega^s$
at time level $t^{n}$, for $q=0,\dots,Q$. Thus, assuming that the
solution is known up to time $t^{n-1}$ and has been converged
using Schwarz iterations at the previous timestep,
$\bphi^{s,n-1,Q}$ represents the solution at time $t^{n-1}$.
With this notation, and assuming a constant timestep size $\dt$
(which is equal for all overlapping grids), we define the Schwarz
update procedure as follows:
\begin{enumerate}[leftmargin=*]

\item Compute the tentative velocity field $\bust$ using
(\ref{eq:vstarsol}) with the solution from $k$ previous
timesteps in each subdomain $\Omega^s, s=1\dots S$:
\begin{eqnarray} \label{eq:vstarsolnn}
\bust^{s,n} &=& -\sum_{j=1}^{k} \beta_j\bu^{s,n-j,Q} + \dt\,\sum_{j=1}^{k} \alpha_{j}
\bffe^{s,n-j,Q},
\end{eqnarray}
where $\bust^{s,n}$ has contributions from the BDF$k$ and EXT$k$
terms. We note that we do not use the superscript $(\,\,)^{q}$
in $\bust^{s,n}$ because it depends only on the solution at
previous timesteps and does not change at each Schwarz
iteration.

\item Use $Q$ simultaneous Schwarz iterations to solve the {\em linear} Stokes subproblems (\ref{eq:prsol})
and (\ref{eq:velsol}) to yield the velocity-pressure pair,
$\bphi^{s,n,q}=[\bu^{s,n,q},p^{s,n,q}]^T$.
\end{enumerate}
\begin{itemize}[leftmargin=0em]
\item Prior to each Schwarz iteration, interpolate the interdomain
boundary data (global process). Since the solution is
known up to $t^{n-1}$, the initial iterate ($q=0$, the {\em predictor step})
uses interdomain boundary data based on $m$th-order extrapolation in time.
The $Q$ subsequent Schwarz
iterations (the {\em corrector steps}) directly interpolate the interdomain
boundary data from the most recent iteration:
\begin{alignat}{3}
\hspace{-0mm}
&q=0:\,\, &&\bhu^{s,n,0} \rvert_{\dO^s_I} = \sum_{j=1}^{m} \textw_{j}\,
\mathcal{I}\bigg(\bu^{r,n-j,Q}\bigg), \label{eq:unsteadystokesschwarz3int}\\
&q=1 \dots Q:\,\, && \bhu^{s,n,q} \rvert_{\dO^s_I} = \mathcal{I}(\bu^{r,n,q-1}), \label{eq:unsteadystokesschwarz3qint}
\end{alignat}
where $m$ is the order of extrapolation for the interdomain
boundary data at $q=0$ iteration, $\textw_j$ are the
corresponding extrapolation weights that are computed using the routines
described in \cite{fornberg1998practical}, and $\mathcal{I}$ is the interpolation operator described in Section \ref{sec:interp}. Note that each Schwarz iteration in singlerate
timestepping effectively requires 1 interpolation, even though the predictor step ($q=0$)
depends on $m$ most recent solutions (e.g., $\bu^{r,n-1,Q}, \bu^{r,n-2,Q}, \bu^{r,n-3,Q}$).
This is because all solutions except the most recent one (e.g., $\bu^{r,n-2,Q}$ and $\bu^{r,n-3,Q}$)
are already available since they must have been interpolated at the predictor step of the previous timestep.
\item Solve the unsteady Stokes problem (locally) in each subdomain:
\begin{alignat}{3}
\hspace{-0mm}
&q=0 \dots Q:\,\, &&{\bf S} \bphi^{s,n,q} &&= \br^{s,n,q}, \,\,\,\,
  \bu^{s,n,q} \rvert_{\dO^s_I} = \bhu^{s,n,q} \rvert_{\dO^s_I} + \btu\rvert_{\dO^s_I},\,\, \label{eq:unsteadystokesschwarz3q}
\end{alignat}
where $\btu\rvert_{\dO^s_I}$ is a flux-based
correction (Section \ref{sec:mass}) required to satisfy the divergence-free constraint,
and
\begin{equation}
\begin{aligned}
\label{eq:unsteadystokesbrnn}
\br^{s,n,q}&=[\br_v^{s,n,q},r_p^{s,n,q}]^T, \\
\br_v^{s,n,q} &= -\nabla{p}^{s,n,q} + \frac{\bust^{s,n}}{\dt}, \\
r_p^{s,n,q} &= -\frac{\nabla \cdot \bust^{s,n}}{\dt} + \frac{1}{Re} \nabla \cdot
\sum_{j=1}^{k} \alpha_{j} (\nabla \times \boldsymbol\omega^{s,n-j,Q}).
\end{aligned}
\end{equation}
In \eqref{eq:unsteadystokesschwarz3q}, we
solve the full unsteady Stokes problem for velocity and pressure
in each subdomain, $\Omega^s$. This unsteady Stokes solve is similar to
the monodomain case \eqref{eq:unsteadystokes}, with the addition of a
Dirichlet condition on the interdomain boundaries ($\dO^s_I$) that is obtained by
interpolating the solution from the overlapping subdomain (prior to each Schwarz iteration)
and imposing a flux-based correction ($\btu\rvert_{\dO^s_I}$).
\end{itemize}
Note that besides the interpolation of the interdomain boundary data,
which is a global process and requires communication between overlapping subdomains,
the unsteady Stokes solve is independent for each subdomain and
utilizes the same matrix-free framework that is used for a monodomain grid
(Section \ref{sec:monoinse}).

The temporal accuracy of this singlerate
timestepping scheme \eqref{eq:unsteadystokesschwarz3q}
is $O(\dt^{\text{min}(m,k)})$. We typically
set $m=k$, unless otherwise stated, and Peet and Fischer
\cite{peet2012} have shown that $Q=0$ is sufficient for $m=1$ and
$Q=1-3$ is sufficient for $m>1$
from a stability point of view. In \cite{mittal2019nonconforming},
we further demonstrated that $Q=0$ is sufficient from an accuracy point of view for
basic statistics (e.g., mean or rms) of turbulent flows. This predictor-corrector
approach has been used in the Schwarz-SEM
framework \cite{merrill2017effect,merrill2019moving,mittal2019direct}
to demonstrate that it maintains the spatial and temporal
convergence of the underlying monodomain SEM framework, and is
effective for solving highly turbulent flow phenomenon in complex
domains using an arbitrary number of overlapping grids.

\subsubsection{Interpolation} \label{sec:interp}
Since overlapping grids rely on interpolation for interdomain boundary
data, the interpolation operator ($\mathcal{I}$) is
of central significance for overlapping Schwarz based methods. In our framework,
$\mathcal{I}$ is effected (in parallel) via \fpt, a scalable high-order
interpolation utility that is part of \emph{gslib} \cite{gslibrepo},
an open-source communication library that readily links with Fortran, C, and
C++ codes.

\fpt\ provides two key functionalities. First, for a given
set of interdomain boundary points that are tagged with the associated subdomain
number $\bx^* = (\bx_1^*, \bx_2^* \dots \bx_b^*)^s$,
\fpt\ determines the computational coordinates of each point. These computational
coordinates ($\bq={r,e,\bxi,p}$) for each point specify the subdomain number $r$
that it overlaps, the element number ($e \in \Omega^r$) in which the
point was found, and
the corresponding reference-space coordinates ($\bxi = (\xi,\eta,\zeta)$)) inside that
element. Since a mesh could be partitioned on to many MPI ranks, \fpt\
also specifies the MPI rank $p$ on which the donor element is located.
For cases where $S=2$, the donor element search is straightforward \cite{noorani2016particle}
because \fpt\
is only concerned with the elements that are not located in the same subdomain
as the sought-point ($e \in \Omega^r, r\neq s$).
In cases where $S>2$, an interdomain boundary point can overlap multiple
subdomains. In these cases, the donor element is chosen from the subdomain that minimizes
the error due to simultaneous Schwarz iterations \cite{mittal2019nonconforming}.
If the nodal positions of all the overlapping grids are fixed in time,
the computational coordinate search needs to be done only at the beginning of the calculation.
Otherwise, the computational coordinate search is done at the beginning of each time-step.

The second key functionality of \fpt\ is that
for a given set of computational coordinates, it can interpolate any scalar
function defined on the spectral element mesh. All the parallel communication
in \fpt\ is handled by \emph{gslib}'s generalized and scalable all-to-all utility, \emph{gs\_crystal},
which is based on the crystal router algorithm of \cite{tufo2001}.
Using \emph{gslib}, \fpt\ has demonstrated excellent scaling in parallel for finding
computational coordinates of a given set of points and interpolating solution in a mondomain mesh \cite{duttaefficient}
and in overlapping meshes \cite{mittal2019nonconforming}.

\subsubsection{Mass-Balance}\label{sec:mass}
While the \emph{gslib}-based interpolation is exponentially convergent,
there can still be mass balance errors arising from the interpolation of the
velocities on the interdomain boundaries. Since the pressure solve is sensitive
to these inconsistencies, we must ensure that for each subdomain $\Omega^s$,
the following compatibility condition for INSE is satisfied at all times:
\begin{eqnarray}
\label{eq:consv1}
\int_{\dO^s} \bu \cdot \bhn &=&0,
\end{eqnarray}
where $\bhn$ represents the outward pointing unit normal vector on $\dO^s$.
In the case of overlapping grids, when a subdomain does not have any natural
boundary conditions ($\dO_N \cap \dO^s=0$), there is a
potential to fail to satisfy (\ref{eq:consv1}) because the interpolated fluxes
on $\dO^s$ may not integrate to zero.

Let $\bhu$ denote the tentative velocity
field defined on $\dO^s$ through prescribed Dirichlet data on $\dO^s_D := \dO^s \cap
\dO_D$ and interpolation on $\dO^s_I$. Assume a correction of the form
\begin{eqnarray}\label{eq:consv2}
\bu\rvert_{\dO^s_I} = \bhu\rvert_{\dO^s_I} + \btu\rvert_{\dO^s_I},
\end{eqnarray}
where $\bu$ is the flux-corrected boundary data on $\dO^s$ and $\btu$ is the
correction required to satisfy (\ref{eq:consv1}). In \cite{mittal2019nonconforming},
we demonstrated that the choice
\begin{eqnarray} \label{eq:corr}
\left. \btu \right|^{}_{\dO^s_I}
&=& \left. \delta \bhn \right|^{}_{\dO^s_I}
\end{eqnarray}
is the $L^2$ minimizer of possible trace-space corrections,
$(\bu-\bhu)$,
that allows (\ref{eq:consv1}) to be satisfied, provided that
\begin{eqnarray} \label{eq:consv3}
\delta
&=&
- \frac{ \int_{\dO^s} \bhu \cdot \bhn \, dA }{ \int_{\dO^s_I} \bhn \cdot \bhn \, dA }
= - \frac{ \int_{\dO^s_D} \bhu \cdot \bhn \, dA + \int_{\dO^s_I} \bhu \cdot \bhn \, dA }{ \int_{\dO^s_I} \bhn \cdot \bhn \, dA }.
\end{eqnarray}
Here, the first term in the numerator ($\int_{\dO^s_D} \bhu \cdot \bhn \, dA$)
is known by definition since it is the prescribed boundary data on
$\dO^s_D$. The second term, $\int_{\dO^s_I} \bhu \cdot \bhn \, dA$,
is based on the interpolated data \eqref{eq:unsteadystokesschwarz3int}-\eqref{eq:unsteadystokesschwarz3qint},
obtained prior to each Schwarz iteration.
Using this flux-correction approach, we modify the interpolated boundary data
for each unsteady Stokes solve \eqref{eq:unsteadystokesschwarz3q}.
Note that unlike interdomain interpolation, which is a global process
and requires communication between all overlapping subdomains, the flux-based
correction is local to each subdomain since it only depends on $\dO^s$.

\section{Methodology for Multirate Timestepping}\label{sec:method}
In this section, we introduce the parallel multirate timestepping scheme
for solving INSE in overlapping subdomains. For simplicity, we first describe our
method in the context of two overlapping subdomains with a fixed timestep ratio
of 2. Next, we describe our method for an arbitrary timestep ratio, since
this is an obvious next step and a scenario that we encounter in most of our applications
(e.g., thermally-buoyant plume in Section \ref{sec:intro} that
is modeled with overlapping domains with a timestep
ratio of 100). Finally, we generalize the MTS scheme to an arbitrary number of
overlapping subdomains with an arbitrary timestep ratio.

\begin{figure}[h!] \begin{center}
$\begin{array}{c}
\includegraphics[width=70mm]{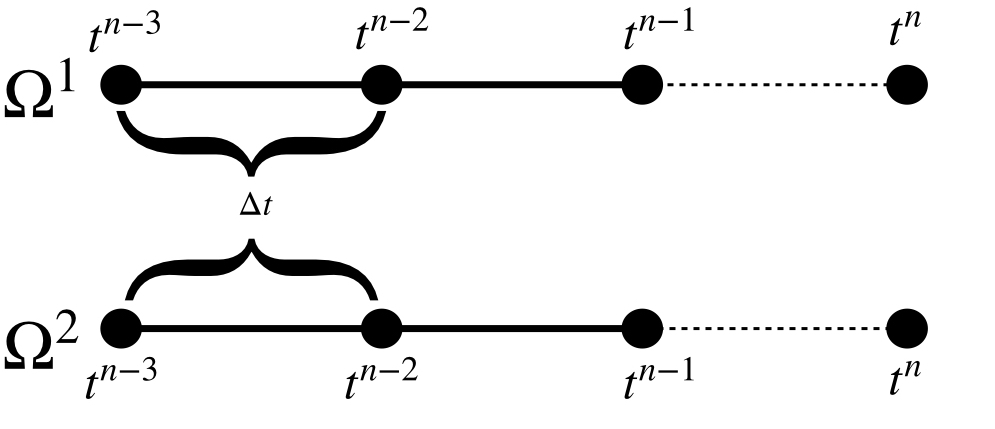} \\
\textrm{(a) Singlerate timestepping.} \\
\includegraphics[width=70mm]{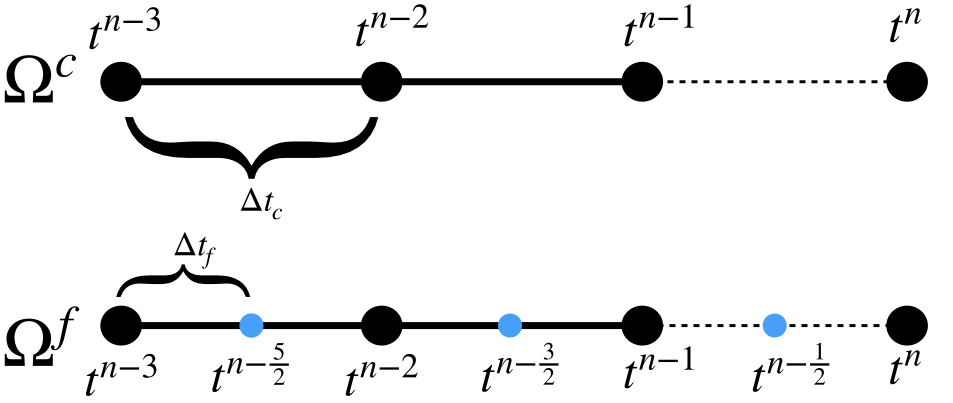} \\
\textrm{(b) Multirate timestepping with timestep ratio $\tsr=2$.}
\end{array}$
\end{center}
\caption{Schematic showing discrete time-levels for singlerate and multirate timestepping ($\tsr=2$).}
\label{fig:schematic}
\end{figure}

In our method, we consider only integer timestep ratios,
\begin{eqnarray}
\tsr := \frac{\dt_c}{\dt_f} \in \mathbb{Z}^{+},
\end{eqnarray}
where $\dt_c$ corresponds to the subdomain ($\Omega^c$) with slower time-scales and $\dt_f$ corresponds to the subdomain ($\Omega^f$) with faster time-scales. Figure \ref{fig:schematic} shows a schematic of the discrete time-levels for
the STS scheme and the MTS scheme with $\tsr=2$. Here, the black circles ($\CIRCLE$)
indicate the timestep levels for both $\Omega^f$ and $\Omega^c$ and the blue
circles (\tikzcircle[cyan,fill=cyan]{2.5pt}) indicate the sub-timestep levels for $\Omega^f$.

\subsection{Multirate Timestepping for $\tsr=2$}
In a slowest- or fastest-first method, assuming the solution is known up to $t^{n-1}$,
the PDE of interest is temporally integrated in either of the domains (e.g., say $\Omega^f$)
to obtain the solution at time $t^n$, which is then used to obtain interdomain
boundary data for advancing the solution in the other domain (e.g., $\Omega^c$).
For a parallel multirate scheme, however, we wish to simultaneously advance
the solution in $\Omega^f$ and $\Omega^c$.  As a result, the interdomain boundary
data is exchanged prior to starting the solution process such that the
unsteady Stokes solve in each subdomain ($\tsr$ sub-timesteps in $\Omega^f$ and
1 timestep in $\Omega^c$) can be completed independently, similar to the
singlerate timestepping (Section \ref{sec:methodschwarz}).
The synchronization time-levels at which the interdomain boundary data is
exchanged are indicated by ($\CIRCLE$) in Fig. \ref{fig:schematic}.

\begin{figure}[t!] \begin{center}
$\begin{array}{c}
\includegraphics[width=100mm]{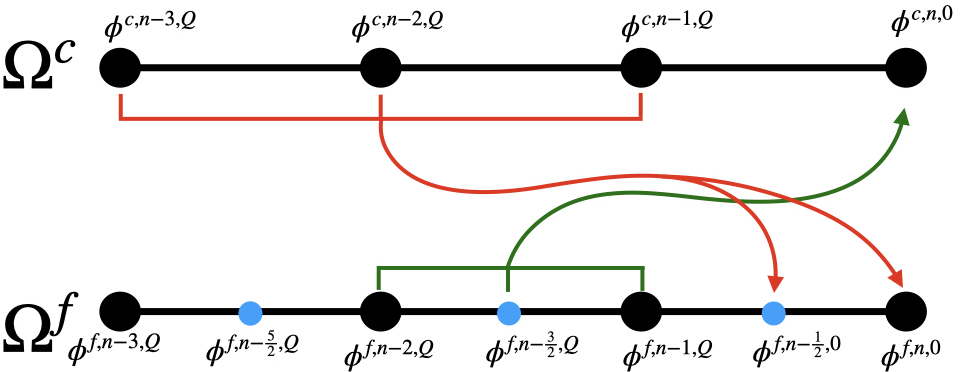}
\end{array}$
\end{center}
\caption{Schematic showing the dependence of the interdomain boundary data for the predictor step.}
\label{fig:schematic_ibd_pred_dep}
\end{figure}

Similar to the singlerate timestepping scheme (e.g., see (\ref{eq:unsteadystokesschwarz3q})),
high-order temporal accuracy is achieved in the multirate setting by extrapolating
the interdomain boundary data obtained from the solution at previous (sub-)
timesteps. For $\tsr=2$, the interdomain boundary data dependency for
the \emph{predictor} step is depicted in Fig. \ref{fig:schematic_ibd_pred_dep}.
For the solutions $\bphi^{f,n-\frac{1}{2},0}$ and
$\bphi^{f,n,0}$ the boundary data is interpolated from the known solutions in $\Omega^c$:
$\bphi^{c,n-1,Q}$, $\bphi^{c,n-2,Q}$, and $\bphi^{c,n-3,Q}$. Simultaneously, the
interdomain boundary data for the solution $\bphi^{c,n,0}$ is
interpolated from the known solutions in $\Omega^f$: $\bphi^{f,n-1,Q}$,
$\bphi^{f,n-\frac{3}{2},Q}$, and $\bphi^{f,n-2,Q}$. This interdomain boundary data exchange occurs
at synchronization time-level $t^{n-1}$,
prior to starting the solution process for times $t^{n-\frac{1}{2}}$ and $t^n$.

\begin{figure}[t!] \begin{center}
$\begin{array}{c}
\includegraphics[width=100mm]{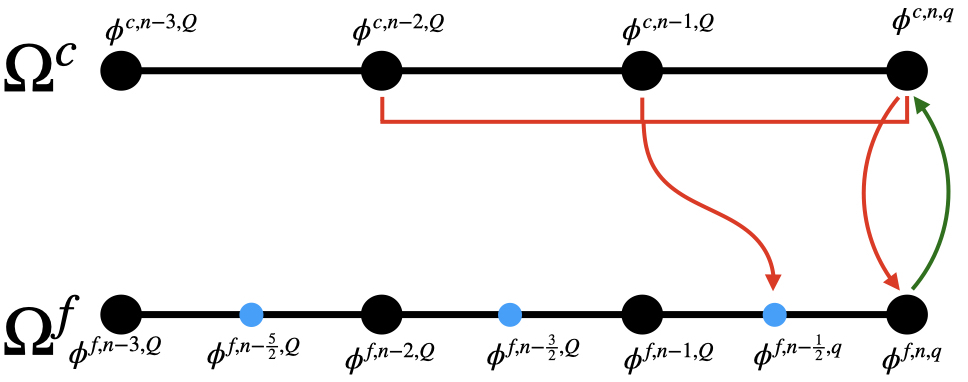}
\end{array}$
\end{center}
\caption{Schematic showing the dependence of the interdomain boundary data for
corrector steps.}
\label{fig:schematic_ibd_cor_dep}
\end{figure}

Once the solution $\bphi^{f,n,0}$ and $\bphi^{c,n,0}$ have been
determined, $Q$ correction iterations are needed to
stabilize the solution if high-order extrapolation is used for interdomain
boundary data during the predictor step. The interdomain boundary data dependency
for the \emph{corrector} steps is depicted in Fig. \ref{fig:schematic_ibd_cor_dep}.
In $\Omega^f$, the interdomain boundary data for $\bphi^{f,n-\frac{1}{2},q}$
comes from the most recent iteration in $\Omega^c$ ($\bphi^{c,n,q-1}$) and
the converged solution at previous timesteps, $\bphi^{c,n-1,Q}$ and
$\bphi^{c,n-2,Q}$. For the solution at time $t^{n}$ ($\bphi^{f,n,q}$ and $\bphi^{c,n,q}$),
the interdomain boundary data only depends on the solution from the most recent
iteration ($\bphi^{c,n,q-1}$ and $\bphi^{f,n,q-1}$).

Using this approach for obtaining interdomain boundary data, we now summarize the multirate timestepping scheme for $\tsr=2$. Recall our notation for singlerate timestepping, $\bphi^{s,n,q}$ denotes the solution $\bphi$ in $\Omega^s$ at the $q$th Schwarz iteration at time $t^n$.

\begin{enumerate}[leftmargin=*]
\item For the predictor step ($q=0$), assuming that the solution is known up to time $t^{n-1}$,
interpolate the interdomain boundary data (global process) followed by
locally computing the tentative velocity field $\bust$ (\ref{eq:vstarsol})
and solving the linear Stokes problem in each subdomain: \\
\begin{itemize}[leftmargin=0em]
\item Interpolate the interdomain boundary data required for each sub-timestep
of $\Omega^f$ and the only timestep of $\Omega^c$:
\begin{eqnarray}
\label{eq:eta2bcintf1}
\bhu^{f,n-\frac{1}{2},0} \rvert_{\dO^f_I} = \mathcal{I}\bigg(\sum_{j=1}^{m} \textw_{1j}\,
\bu^{c,n-j,Q}\bigg), \\
\label{eq:eta2bcintf2}
\bhu^{f,n,0} \rvert_{\dO^f_I} = \mathcal{I}\bigg(\sum_{j=1}^{m} \textw_{2j}\,
\bu^{c,n-j,Q}\bigg), \\
\label{eq:eta2bcintc1}
\bhu^{c,n,0} \rvert_{\dO^c_I} = \mathcal{I}\bigg(\sum_{j=1}^{m} \textw_{1j}\,
\bu^{f,n-\frac{j+1}{2},Q}\bigg).
\end{eqnarray}
Here, equations \eqref{eq:eta2bcintf1} and \eqref{eq:eta2bcintf2} refer to
the interdomain boundary data required for the first and second sub-timestep,
respectively, of $\Omega^f$, and
\eqref{eq:eta2bcintc1} refers to the interdomain boundary data required for the
only timestep of $\Omega^c$. In \eqref{eq:eta2bcintf1}-\eqref{eq:eta2bcintc1},
$\gamma_{ij}$ refers to the coefficients used to extrapolate the interdomain
boundary data at the $i$th sub-timestep, which are computed based on the time-levels
(e.g., $t^{n-1}$, $t^{n-\frac{3}{2}}$, etc.)
and the order of extrapolation $m$ such that the desired temporal
accuracy is maintained, using the routines described in \cite{fornberg1998practical}.
\item Unsteady Stokes solve for the first sub-timestep in $\Omega^f$:
\begin{subequations}\label{eq:usmulti2fp1}
\begin{alignat}{2}
&\hspace{-1em}\bust^{f,n-\frac{1}{2}} = -\sum_{j=1}^{k} \beta_j\bu^{f,n-\frac{j+1}{2},Q} +
\dt_f\,\sum_{j=1}^{k} \alpha_{j} \bffe^{f,n-\frac{j+1}{2},Q}, \\
&\hspace{-1em}{\bf S} \bphi^{f,n-\frac{1}{2},0} = \br^{f,n-\frac{1}{2},0}, \,\,\,\,\,\,
\bu^{f,n-\frac{1}{2},0} \rvert_{\dO^f_I} = \bhu^{f,n-\frac{1}{2},0} \rvert_{\dO^f_I} + \delta \bhn \rvert_{\dO^f_I} .\,\,\,\,\,\,\,\,
\end{alignat}
\end{subequations}

\item Unsteady Stokes solve for the second sub-timestep in $\Omega^f$:
\begin{subequations}
\label{eq:usmulti2fp2}
\begin{alignat}{2}
&\hspace{-1em}\bust^{f,n} = -\sum_{j=1}^{k} \beta_j\bu^{f,n-\frac{j}{2},Q} +
\dt_f\,\sum_{j=1}^{k} \alpha_{j} \bffe^{f,n-\frac{j}{2},Q},   \\
&\hspace{-1em}{\bf S} \bphi^{f,n,0} = \br^{f,n,0}, \,\,\,\,\,\,
\bu^{f,n,0} \rvert_{\dO^f_I} = \bhu^{f,n,0} \rvert_{\dO^f_I} + \delta \bhn \rvert_{\dO^f_I}.
\end{alignat}
\end{subequations}

\item Unsteady Stokes solve for the only timestep in $\Omega^c$:
\begin{subequations}
\label{eq:usmulti2cp}
\begin{alignat}{2}
&\bust^{c,n} = -\sum_{j=1}^{k} \beta_j\bu^{c,n-j,Q} +
\dt_c\,\sum_{j=1}^{k} \alpha_{j} \bffe^{c,n-j,Q},   \\
&{\bf S} \bphi^{c,n,0} = \br^{c,n,0}, \,\,\,\,\,\,
 \bu^{c,n,0} \rvert_{\dO^c_I} = \bhu^{c,n,0} \rvert_{\dO^c_I} + \delta \bhn \rvert_{\dO^c_I}.
\end{alignat}
\end{subequations}

\end{itemize}

In (\ref{eq:usmulti2fp1}), we first compute the tentative velocity field,
similar to (\ref{eq:vstarsolnn}) for singlerate timestepping, and then solve
for $\bphi^{f,n-\frac{1}{2},0}$ using the flux-corrected interdomain boundary
data (Section \ref{sec:mass}). Note that computing $\delta$ for flux-correction
through \eqref{eq:consv3} requires the interpolated interdomain boundary data
\eqref{eq:eta2bcintf1}-\eqref{eq:eta2bcintc1} and the prescribed Dirichlet
data on $\dO^s_D$ (which is known by definition). Additionally, $\br^{f,n-\frac{1}{2},q}$ is
unchanged from (\ref{eq:unsteadystokesbrnn}) for singlerate timestepping scheme.

Once $\bphi^{f,n-\frac{1}{2},0}$ is determined, $\bphi^{f,n,0}$ is computed using (\ref{eq:usmulti2fp2}), which completes the predictor step for advancing the solution of INSE in $\Omega^f$. Parallel to (\ref{eq:usmulti2fp1}) and (\ref{eq:usmulti2fp2}), (\ref{eq:usmulti2cp}) is used for the solution at time $t^n$ in $\Omega^c$.

\item Once the predictor step, $q=0$, is complete, $q=1 \dots Q$ corrector iterations are done to improve the accuracy of the solution and stabilize the method:

\begin{itemize}[leftmargin=0em]
\item Interpolate the interdomain boundary data:
\begin{eqnarray}
\label{eq:eta2bcintcf1}
\bhu^{f,n-\frac{1}{2},q} \rvert_{\dO^f_I} &=& \mathcal{I}\bigg(\tintw_{11}\bu^{c,n,q-1}
+\tintw_{12}\bu^{c,n-1,Q}+\tintw_{13}\bu^{c,n-2,Q} \bigg), \\
\label{eq:eta2bcintcf2}
\bhu^{f,n,q} \rvert_{\dO^f_I} &=& \mathcal{I}\bigg(\bu^{c,n,q-1}\bigg), \\
\label{eq:eta2bcintcc1}
\bhu^{c,n,q} \rvert_{\dO^c_I} &=& \mathcal{I}\bigg(\bu^{f,n,q-1}\bigg).
\end{eqnarray}
In \eqref{eq:eta2bcintcf1}, the coefficients for the temporal interpolation
represented by $\tintw_{1j}$ are computed assuming linear
interpolation when $m=1$ or $2$, and quadratic interpolation when $m=3$.
This approach ensures that the desired temporal accuracy $\mathcal{O}(\dt^m)$ is
maintained. In \eqref{eq:eta2bcintcf2} and \eqref{eq:eta2bcintcc1}, the interdomain
boundary data is directly spatially interpolated without the need
for any temporal interpolation.
\color{black}
\item Unsteady Stokes solve for the first sub-timestep in $\Omega^f$:
\begin{subequations}
\label{eq:usmulti2fc1}
\begin{alignat}{2}
&\hspace{-1em}\bust^{f,n-\frac{1}{2}} = -\sum_{j=1}^{k} \beta_j\bu^{f,n-\frac{j+1}{2},Q} +
\dt_f\,\sum_{j=1}^{k} \alpha_{j} \bffe^{f,n-\frac{j+1}{2},Q},   \\
&\hspace{-1em}{\bf S} \bphi^{f,n-\frac{1}{2},q} = \br^{f,n-\frac{1}{2},q}, \,\,\,\,\,\,
\bu^{f,n-\frac{1}{2},q} \rvert_{\dO^f_I} = \bhu^{f,n-\frac{1}{2},q} \rvert_{\dO^f_I} + \delta \bhn \rvert_{\dO^f_I}.\,\,
\end{alignat}
\end{subequations}

\item Unsteady Stokes solve for the second sub-timestep in $\Omega^f$:
\begin{subequations}
\label{eq:usmulti2fc2}
\begin{alignat}{2}
&\hspace{-1em}\bust^{f,n} = -\sum_{j=1}^{k} \beta_j\bu^{f,n-\frac{j}{2},Q} +
\dt_f\,\sum_{j=1}^{k} \alpha_{j} \bffe^{f,n-\frac{j}{2},Q},   \\
&\hspace{-1em}{\bf S} \bphi^{f,n,q} = \br^{f,n,q}, \,\,\,\,\,\,
\bu^{f,n,q} \rvert_{\dO^f_I} = \bhu^{f,n,q} \rvert_{\dO^f_I} + \delta \bhn \rvert_{\dO^f_I}.\,\,
\end{alignat}
\end{subequations}

\item Unsteady Stokes solve for the only timestep in $\Omega^c$:
\begin{equation}
\label{eq:usmulti2cc}
\begin{aligned}
{\bf S} \bphi^{c,n,q} = \br^{c,n,q}, \,\,\,\,\,\,
  \bu^{c,n,q} \rvert_{\dO^c_I} = \bhu^{c,n,q} \rvert_{\dO^c_I} + \delta \bhn \rvert_{\dO^c_I}.\,\,
\end{aligned}
\end{equation}

\end{itemize}

In (\ref{eq:usmulti2fc1}), we compute $\bphi^{f,n-\frac{1}{2},q}$ using the
flux-corrected interdomain boundary data, followed by (\ref{eq:usmulti2fc2})
to determine $\bphi^{f,n,q}$ in $\Omega^f$. Similarly, $\bphi^{c,n,q}$ is
(concurrently) computed using (\ref{eq:usmulti2cc}).

We note that in the singlerate timestepping scheme, the tentative velocity field
($\bust$) was computed only once for the $Q$ corrector iterations (\ref{eq:vstarsolnn}).
In contrast, we recompute the tentative velocity field in (\ref{eq:usmulti2fc1})--(\ref{eq:usmulti2fc2})
at each corrector iteration for $\Omega^f$, because the solution process spans
$\tsr$ multiple sub-timesteps.  Saving $\bust$ for each of the $\tsr$
sub-timesteps is not a scalable approach (e.g., $\tsr=100$ will require us to
save tentative velocity field for 100 sub-timesteps).
Using \eqref{eq:eta2bcintcf1}-(\ref{eq:usmulti2cc}), $Q$ simultaneous Schwarz iterations
determine the solution in $\Omega^f$ and $\Omega^c$ at time $t^n$.
\end{enumerate}

\subsection{Multirate Timestepping for Arbitrary $\tsr$}
\begin{figure}[t!] \begin{center}
$\begin{array}{c}
\includegraphics[width=100mm]{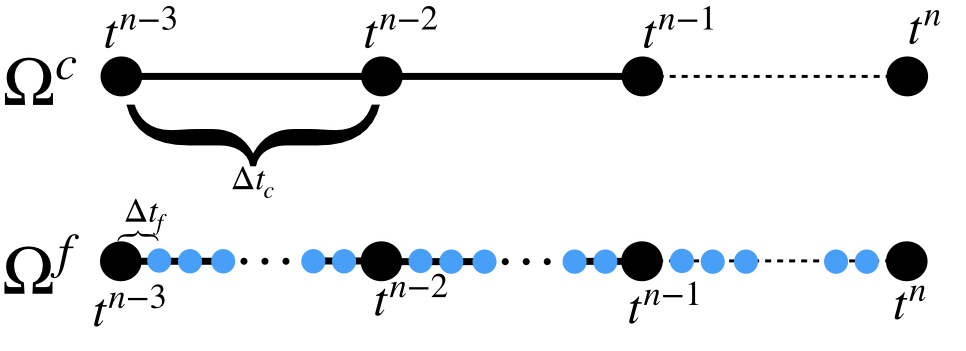}
\end{array}$
\end{center}
\caption{Schematic showing discrete time-levels for the multirate timestepping with an arbitrary timestep ratio.}
\label{fig:arbitraryschematic}
\end{figure}

From the preceding discussion, we can anticipate that the generalization
of this multirate scheme will require $\tsr$ sub-timesteps in $\Omega^f$
and only one timestep in $\Omega^c$.
Figure \ref{fig:arbitraryschematic} shows a schematic with time-levels for an arbitrary timestep ratio.
Similar to \eqref{eq:eta2bcintf1}-(\ref{eq:usmulti2cc}),
the timestepping strategy for arbitrary (integer)
$\tsr$ is,

\begin{enumerate}[leftmargin=*]

\item Compute the tentative velocity field and solve the linear Stokes problem in each subdomain for the predictor step ($q=0$).

\begin{itemize}[leftmargin=0em]
\item Interpolate the interdomain boundary data for $i=1\dots\tsr$ sub-timesteps of
$\Omega^f$ and 1 timestep of $\Omega^c$:
\begin{eqnarray}
\label{eq:etamanybcintfp1}
\bhu^{f,n-1+\frac{i}{\tsr},0} \rvert_{\dO^f_I} &=& \mathcal{I}\bigg(\sum_{j=1}^{m} \textw_{ij}\,
\bu^{c,n-j,Q}\bigg),\qquad i=1\dots\tsr\\
\label{eq:etamanybcintcp1}
\bhu^{c,n,0} \rvert_{\dO^c_I} &=& \mathcal{I}\bigg(\sum_{j=1}^{m} \textw_{1j}\,
\bu^{f,n-1-\frac{j-1}{\tsr},Q}\bigg).\,\,
\end{eqnarray}
\color{black}
\item Unsteady Stokes solve for the $i=1\dots\tsr$ sub-timesteps of
$\Omega^f$:
\begin{subequations}\label{eq:usmultietafp}
\begin{alignat}{3}
&\hspace{-5em}\bust^{f,n-1+\frac{i}{\tsr}} && = && -\sum_{j=1}^{k}
\beta_j\bu^{f,n-1-\frac{j-i}{\tsr},Q}\,\, + \dt_f\,\sum_{j=1}^{k} \alpha_{j}
\bffe^{f,n-1-\frac{j-i}{\tsr},Q},   \\
&\hspace{-5em}{\bf S} \bphi^{f,n-1+\frac{i}{\tsr},0} && = && \br^{f,n-1+\frac{i}{\tsr},0}, \,\,\,\,\,\,
\bu^{f,n-1+\frac{i}{\tsr},0} \rvert_{\dO^f_I} = \bhu^{f,n-1+\frac{i}{\tsr},0} \rvert_{\dO^f_I} + \delta \bhn \rvert_{\dO^f_I}.
\end{alignat}
\end{subequations}

\item Unsteady Stokes solve for the only timestep of $\Omega^c$:
\begin{subequations}\label{eq:usmultietacp}
\begin{alignat}{2}
&\bust^{c,n} = -\sum_{j=1}^{k} \beta_j\bu^{c,n-j,Q} +
\dt_c\,\sum_{j=1}^{k} \alpha_{j} \bffe^{c,n-j,Q},   \\
&{\bf S} \bphi^{c,n,0} = \br^{c,n,0}, \,\,\,\,\,\,
\bu^{c,n,0} \rvert_{\dO^c_I} = \bhu^{c,n,0} \rvert_{\dO^c_I} + \delta \bhn \rvert_{\dO^c_I}.
\end{alignat}
\end{subequations}
\end{itemize}

In (\ref{eq:usmultietafp}), we compute the sub-timestep solution for $\Omega^f$, sequentially from $i=1 \dots \tsr$, and in (\ref{eq:usmultietacp}), we (concurrently) compute the solution in $\Omega^c$ at time $t^n$.

\item Once the predictor step is complete, $q=1\dots Q$ corrector iterations
are done.

\begin{itemize}[leftmargin=0em]
\item Interpolate the interdomain boundary data:
\begin{eqnarray}
\label{eq:etamanybcintfc1}
\bhu^{f,n-1+\frac{i}{\tsr},q} \rvert_{\dO^f_I} &=& \mathcal{I}\bigg(\tintw_{i1}\bu^{c,n,q-1} + \tintw_{i2}\bu^{c,n-1,Q} +
\tintw_{i3}\bu^{c,n-2,Q}\bigg),\qquad i=1\dots\tsr,\\
\label{eq:etamanybcintcc1}
\bhu^{c,n,q} \rvert_{\dO^c_I} &=& \mathcal{I}\bigg( \bu^{f,n,q-1}\bigg).\,\,
\end{eqnarray}
\color{black}
\item Unsteady Stokes solve for the $i=1\dots\tsr$ sub-timesteps of $\Omega^f$:
\begin{subequations}
\label{eq:usmultietafc}
\begin{alignat}{3}
&\hspace{-5em}\bust^{f,n-1+\frac{i}{\tsr}} && = && -\sum_{j=1}^{k}
\beta_j\bu^{f,n-1-\frac{j-i}{\tsr},Q} \,\,+ \dt_f\,\sum_{j=1}^{k} \alpha_{j}
\bffe^{f,n-1-\frac{j-i}{\tsr},Q},   \\
&\hspace{-5em}{\bf S} \bphi^{f,n-1+\frac{i}{\tsr},q} && = &&\br^{f,n-1+\frac{i}{\tsr},q},
\bu^{f,n-1+\frac{i}{\tsr},q} \rvert_{\dO^f_I} = \bhu^{f,n-1+\frac{i}{\tsr},q} \rvert_{\dO^f_I} + \delta \bhn \rvert_{\dO^f_I}.
\end{alignat}
\end{subequations}

\item Unsteady Stokes solve for the only timestep of $\Omega^c$:
\begin{equation}
\begin{aligned}
\label{eq:usmultietacc}
{\bf S} \bphi^{c,n,q} = \br^{c,n,q}, \,\,\,\,\,\,
\bu^{c,n,q} \rvert_{\dO^c_I} = \bhu^{c,n,q} \rvert_{\dO^c_I} + \delta \bhn \rvert_{\dO^c_I}.\,\,
\end{aligned}
\end{equation}
\end{itemize}

\end{enumerate}
Using the high-order multirate timestepping strategy in \eqref{eq:etamanybcintfp1}-(\ref{eq:usmultietacc}),
the solution to the incompressible Navier-Stokes equations can be advanced in two overlapping
grids for an arbitrary timestep ratio.

\subsection{Multirate Timestepping in $S>2$ Overlapping Domains}
\begin{figure}[h!] \begin{center}
$\begin{array}{c}
\includegraphics[width=100mm]{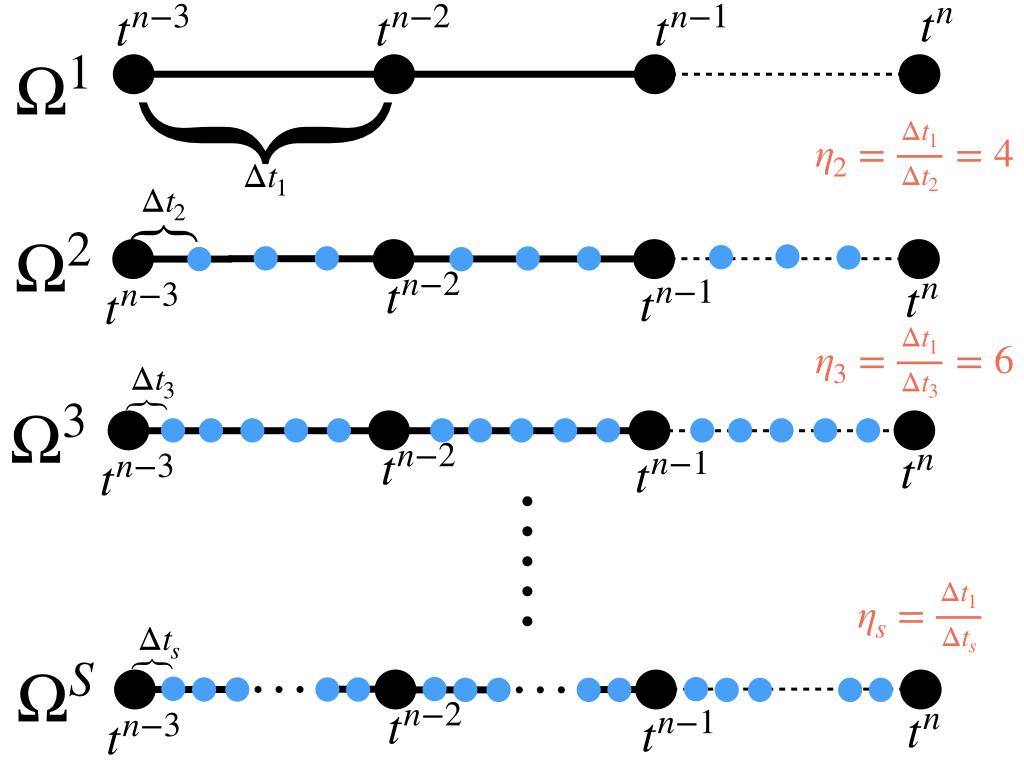}
\end{array}$
\end{center}
\caption{Schematic showing discrete time-levels for the multirate timestepping in an arbitrary number of subdomains.}
\label{fig:arbitrarydomainschematic}
\end{figure}

With the multirate timestepping scheme that we have described, it is straightforward to scale this method to an arbitrary number of domains. Figure \ref{fig:arbitrarydomainschematic} shows an example of a schematic with time-levels for MTS in $s=1\dots S$ subdomains.

For notational purposes, we will use $\Omega^c$ to represent the subdomain with
slowest time-scales.
With each subdomain, we associate the timestep ratio with respect to the timestep size
of $\Omega^c$:
\begin{eqnarray}
\label{eq:tsrmany}
\tsr_{s} := \frac{\dt_c}{\dt_s},
\end{eqnarray}
such that $\tsr_{c}=1$ and $\tsr_{s}>1$ for $s \neq c$. For the example in Fig. \ref{fig:arbitrarydomainschematic}, $c=1$ and the timestep ratios for different subdomains are $\tsr_{1}=1$, $\tsr_{2}=\dt_1/\dt_2=4$ and $\tsr_{3}=\dt_1/\dt_3=6$.

Using \eqref{eq:tsrmany} and assuming that the interdomain boundary data for
points on $\dO_I^s$ is interpolated from $\Omega^r$, the MTS for
arbitrary number of subdomains ($s=1 \dots S$) is:

\begin{enumerate}[leftmargin=*]
\item Solve the unsteady Stokes problem for $i=1\dots\tsr_{s}$ sub-timesteps.
\begin{itemize}[leftmargin=0em]
\item Interpolate the interdomain boundary data (Section \ref{sec:interp}):
\begin{eqnarray}
\label{eq:etamanysbcintfp1}
\bhu^{n-1+\frac{i}{\tsr_{s}}} \rvert_{\dO^s_I} = \mathcal{I}\bigg(\sum_{j=1}^{m} \textw_{ij}\,
\bu^{r,n-1-\frac{j-1}{\tsr_{r}},Q}\bigg)
\end{eqnarray}\color{black}
\item Unsteady Stokes solve in each subdomain (Section \ref{sec:monoinse})
for the $i=1\dots\tsr_{s}$ sub-timesteps during the predictor stage with
flux-corrected boundary data on $\dO^s_I$ (Section \ref{sec:mass}):
\begin{subequations}\label{eq:usmultietamfp}
\begin{alignat}{3}
&\hspace{-5em}\bust^{s,n-1+\frac{i}{\tsr_{s}}} && = && -\sum_{j=1}^{k}
\beta_j\bu^{s,n-1-\frac{j-i}{\tsr_{s}},Q}\,\, + \dt_s\,\sum_{j=1}^{k} \alpha_{j}
\bffe^{s,n-1-\frac{j-i}{\tsr_{s}},Q},   \\
&\hspace{-5em}{\bf S} \bphi^{s,n-1+\frac{i}{\tsr_s},0} && = && \br^{s,n-1+\frac{i}{\tsr_s},0},\,\,\,
\bu^{n-1+\frac{i}{\tsr_s}} \rvert_{\dO^s_I} = \bhu^{n-1+\frac{i}{\tsr_s}} \rvert_{\dO^s_I} +
\delta \bhn \rvert_{\dO^s_I}
\end{alignat}
\end{subequations}
\end{itemize}
\item Once the predictor step is complete, do $q=1\dots Q$ corrector iterations.
\begin{itemize}[leftmargin=0em]
\item Interpolate the interdomain boundary data (Section \ref{sec:interp}):
\begin{eqnarray}
\bhu^{n-1+\frac{i}{\tsr_{s}}} \rvert_{\dO^s_I} =
\mathcal{I}\bigg(\tintw_{i1}\bu^{r,n,q-1} + \tintw_{i2}\bu^{r,n-1,Q} +
\tintw_{i3}\bu^{c,n-1-\frac{1}{\tsr_{r}},Q}\bigg)
\end{eqnarray}\color{black}
\item Unsteady Stokes solve in each subdomain (Section \ref{sec:monoinse})
for the $i=1\dots\tsr_{s}$ sub-timesteps during each of the $Q$ corrector
iterations. Each unsteady Stokes solve uses the flux-corrected boundary data
on $\dO^s_I$ (Section \ref{sec:mass}):
\begin{subequations}
\label{eq:usmultietamcc}
\begin{alignat}{3}
&\hspace{-5em}\bust^{s,n-1+\frac{i}{\tsr_{s}}} && = && -\sum_{j=1}^{k}
\beta_j\bu^{s,n-1-\frac{j-i}{\tsr_{s}},Q} \,\,+ \dt_s\,\sum_{j=1}^{k} \alpha_{j}
\bffe^{s,n-1-\frac{j-i}{\tsr_{s}},Q},   \\
&\hspace{-5em}{\bf S} \bphi^{s,n-1+\frac{i}{\tsr_{s}},q} && = &&\br^{s,n-1+\frac{i}{\tsr_{s}},q},\,\,\,
\bu^{n-1+\frac{i}{\tsr_{s}}} \rvert_{\dO^s_I} = \bhu^{n-1+\frac{i}{\tsr_{s}}} \rvert_{\dO^s_I} + \delta \bhn \rvert_{\dO^s_I}.
\end{alignat}
\end{subequations}
\end{itemize}

\end{enumerate}
\color{black}

Equations (\ref{eq:etamanysbcintfp1})-(\ref{eq:usmultietamcc}) describe the
MTS method for solving the INSE in an arbitrary number of overlapping grids
with an arbitrary integer timestep ratio.
From an implementation perspective, since different interdomain boundary points
in a grid can overlap different grids, the coefficients $\textw_{ij}$ and
$\tintw_{ij}$ are computed for each point based on the timestep size of the
donor subdomain ($\Omega^r$). Note that for cases where all the meshes are fixed
and the timestep size is constant in each subdomain,
these coefficients thus need to be computed only at the beginning of the calculation.

\begin{figure}[t!] \begin{center}
$\begin{array}{c}
\includegraphics[width=80mm]{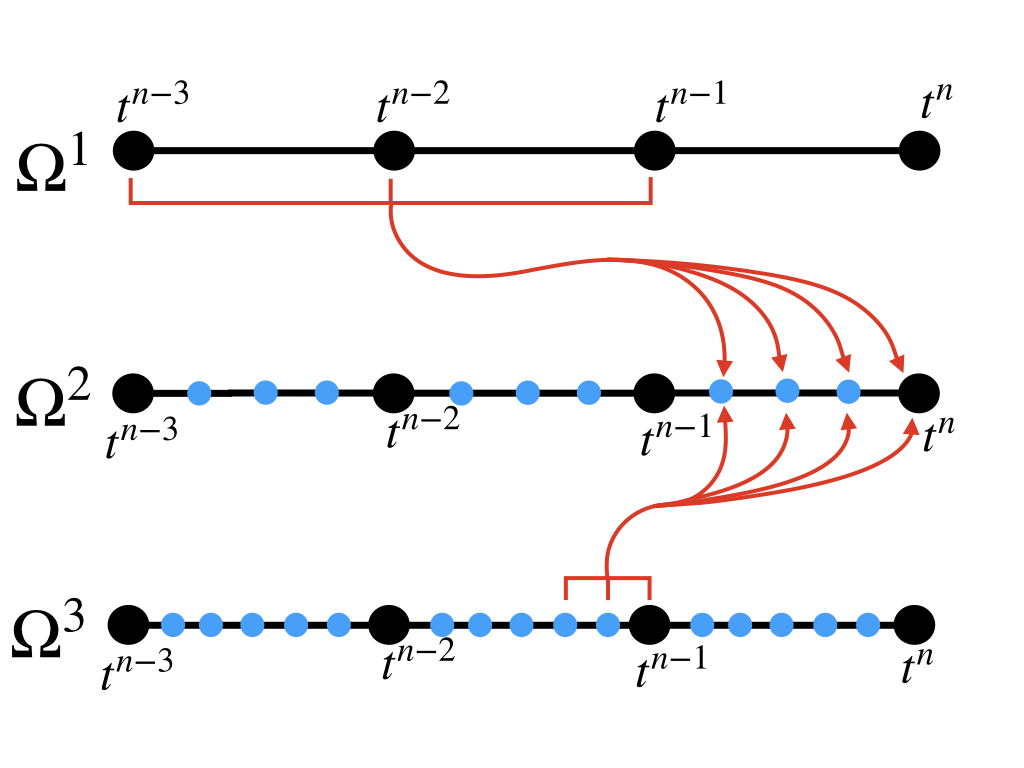} \\
\textrm{(a) Interdomain boundary data dependency for $\Omega^2$ at the predictor step.} \\
\includegraphics[width=80mm]{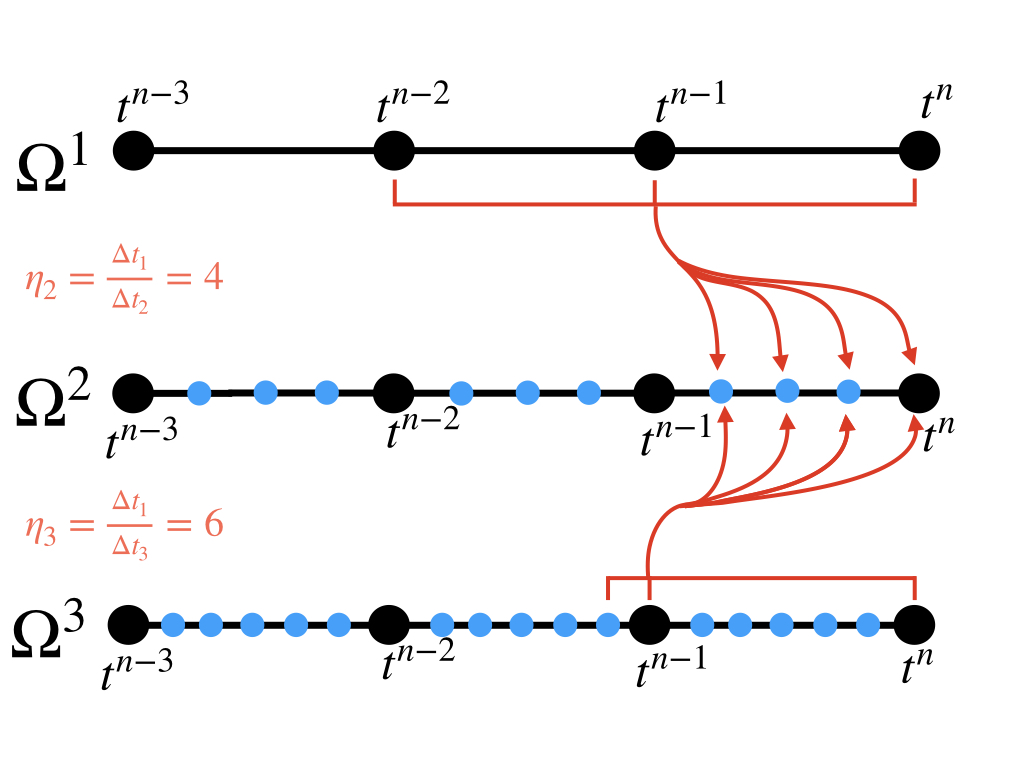} \\
\textrm{(b) Interdomain boundary data dependency for $\Omega^2$ at the corrector step.}
\end{array}$
\end{center}
\caption{Schematic showing interdomain boundary data dependency for $\Omega^2$ from $\Omega^1$ and $\Omega^3$.}
\label{fig:arbitrarydomainschematicpredcor}
\end{figure}

Figure \ref{fig:arbitrarydomainschematicpredcor} shows an example of the interdomain
boundary data dependency for the schematic shown in Fig. \ref{fig:arbitrarydomainschematic}.
Here, we assume that the gridpoints on $\dO_I^2$ overlap
$\Omega^1$ or $\Omega^3$. Assuming that the solution is
know up to time $t^{n-1}$, the boundary data for $\dO_I^2$ is extrapolated for the
predictor step using $\bphi^{r,n-1,Q}$, $\bphi^{r,n-1-\frac{1}{\eta_r},Q}$, and $\bphi^{r,n-1-\frac{2}{\eta_r},Q}$, where $r=1$ or 2.
Similarly,
the boundary data for $\dO_I^2$ for the corrector steps ($q=1\dots Q$) is extrapolated from
the most recent Schwarz iteration $\bphi^{r,n,q-1}$ and the
converged solutions $\bphi^{r,n-1,Q}$ and $\bphi^{r,n-1-\frac{1}{\eta_r},Q}$.

We note that unlike the singlerate timestepping scheme where only 1 interpolation
is required at each predictor and corrector iteration, the multirate timestepping scheme
requires $m$ interpolations at the beginning of each
predictor step and 1 interpolation at the beginning of each corrector iteration.
Thus, for $m$-th order temporal accuracy with an example corresponding to timestep ratio $\tsr$,
the STS scheme requires a total
of $(Q+1)\tsr$ interpolations ($Q+1$ at each sub-timestep)
and the MTS scheme requires a total of $m+Q$ interpolations. Consequently,
the MTS typically requires fewer interpolations in comparison to the STS
scheme. In Section \ref{sec:timinigmulti}, we will use the
example of a thermally-buoyant plume to compare the total time to solution between
the STS and MTS scheme with $\tsr=5$.

\subsection{Stability Considerations}
An underlying assumption of the MTS scheme is that each subdomain has a timestep size
that satisfies its CFL stability criterion \eqref{eq:cflmono}. This requirement
ensures \emph{intradomain} stability, i.e., the unsteady Stokes solve for
time-advancing the solution of the INSE is stable in each subdomain.
\emph{Interdomain} stability, however, is similar to the singlerate timestepping
scheme and depends on the order of extrapolation ($m$) used for interdomain
boundary data and the number of Schwarz iterations ($Q$) used at each timestep.

Peet and Fischer \cite{peet2012} have analyzed the stability of the singlerate
timestepping scheme using an FD-based framework to show than $Q=1-3$ is sufficient
from a stability and accuracy point of view when $m>1$.
We have extended their method to analyze the stability of the MTS scheme for
$S=2$ \cite{mittal2020stability}.
There are two important results that have come forth from this analysis. First,
we note that the MTS scheme requires at-least one Schwarz iteration for stability
when $m>1$, depending on the timestep ratio. For $m=1$, however, $Q=0$ is
sufficient for stability regardless of the timestep ratio. Second,
we observe that for STS, odd values of $Q$ are more stable than even values of
$Q$. In contrast, for MTS with $\tsr=2$, even-$Q$ is more stable than odd-$Q$.
This odd-even
stability pattern goes away for large timestep ratios ($\tsr \geq 4$). The
results that we have observed in our stability analysis are similar to observations
about predictor-corrector
methods by Stetter \cite{stetter1968} and Love et al. \cite{love2009stability}.

In Section \ref{sec:plume}, we will demonstrate that $Q=0,\,m=1$ is sufficient from an accuracy point of view when the subdomains overlap away from the region of interest. This observation is in agreement with the results in \cite{mittal2019nonconforming} that show that for the STS-based Schwarz-SEM framework, the noniterated case ($Q=0$) provides a fast and sufficiently accurate pathway for basic statistics (e.g., mean and rms) of turbulence in complex domains.

\section{Results     }\label{sec:apps}
In this section, we demonstrate the effectiveness of the MTS-based strategy for
solving the incompressible Navier-Stokes equations with two different
examples.  In the first example, we use a problem with a known exact solution
to demonstrate the spatial and temporal convergence of the MTS method. Next, we use this method to
model a buoyant thermal plume in a stratified environment, where the
INSE is solved on overlapping grids with $\tsr=100$.

\subsection{Exact Solution for Decaying Vortices} \label{sec:eddy}
Our first example is due to  Walsh \cite{walsh1992eddy},  who derived a family of exact
Navier-Stokes eigenfunctions that can be used to test spatial and temporal
convergence of discretizations of the INSE.
The eigenfunctions are linear combinations of
$\cos(px) \cos(qy),$ $\sin(px) \cos(qy),$ $\cos(px) \sin(qy),$ and $\sin(px) \sin(qy)$,
for all integer pairs ($p, q$) satisfying $\lambda = -(p^2+q^2)$.  Taking as an initial
condition the eigenfunction $\hat{\bu} = (-{\psi}_y,{\psi}_x)$, a solution to the INSE is
$\bu = e^{\nu \lambda t} \hat{\bu}(\bx)$. Here, $\psi$ is the streamfunction resulting
from the linear combinations of eigenfunctions. Interesting long-time solutions can
be realized by adding a relatively high-speed mean flow $\bu_0$ to the eigenfunction,
in which case the solution is $\bu_{exact} = e^{\nu \lambda t} \hat{\bu}[\bx-\bu_0t]$,
where the brackets imply that the argument is modulo $2\pi$ in $x$ and $y$. As a
result, this problem lets us test our algorithm in the advection-dominated
limit.  (The alternative of simply decreasing $\nu$ can yield to chaotic
solutions because the exact eigenfunctions are not stable solutions to
the INSE at elevated Reynolds numbers.)

\begin{figure}[tbh]
\begin{center}
$\begin{array}{cc}
\includegraphics[height=45mm]{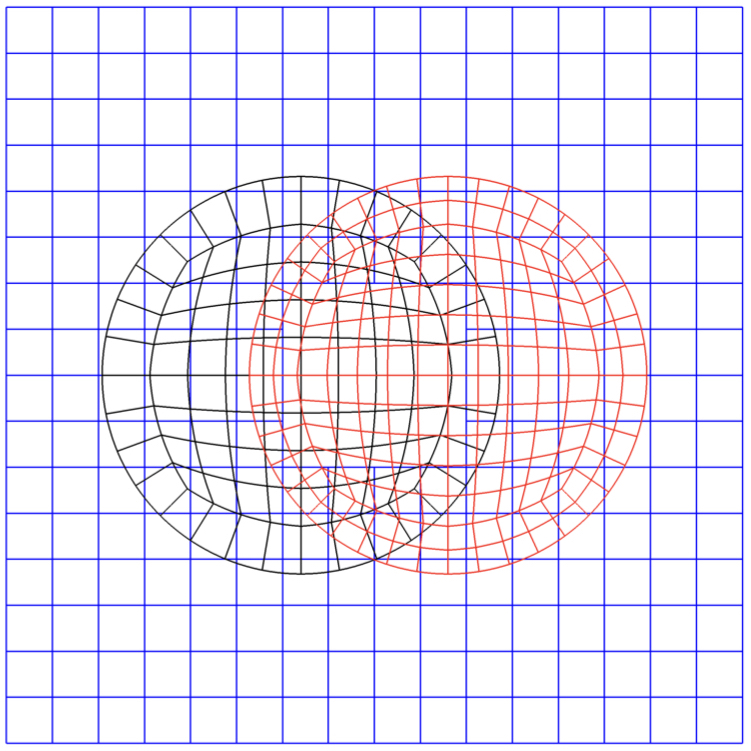} &
\includegraphics[height=45mm]{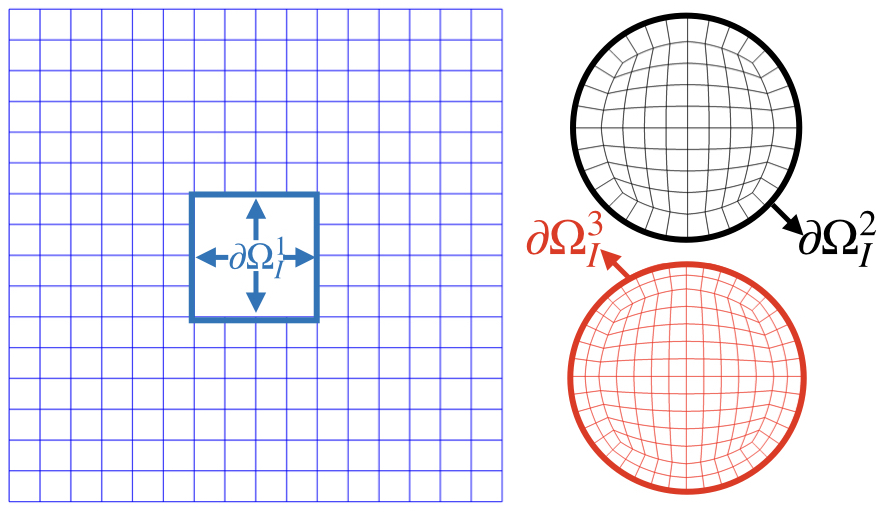} \\
\textrm{(a) } & \textrm{(b) } \\
\includegraphics[height=45mm]{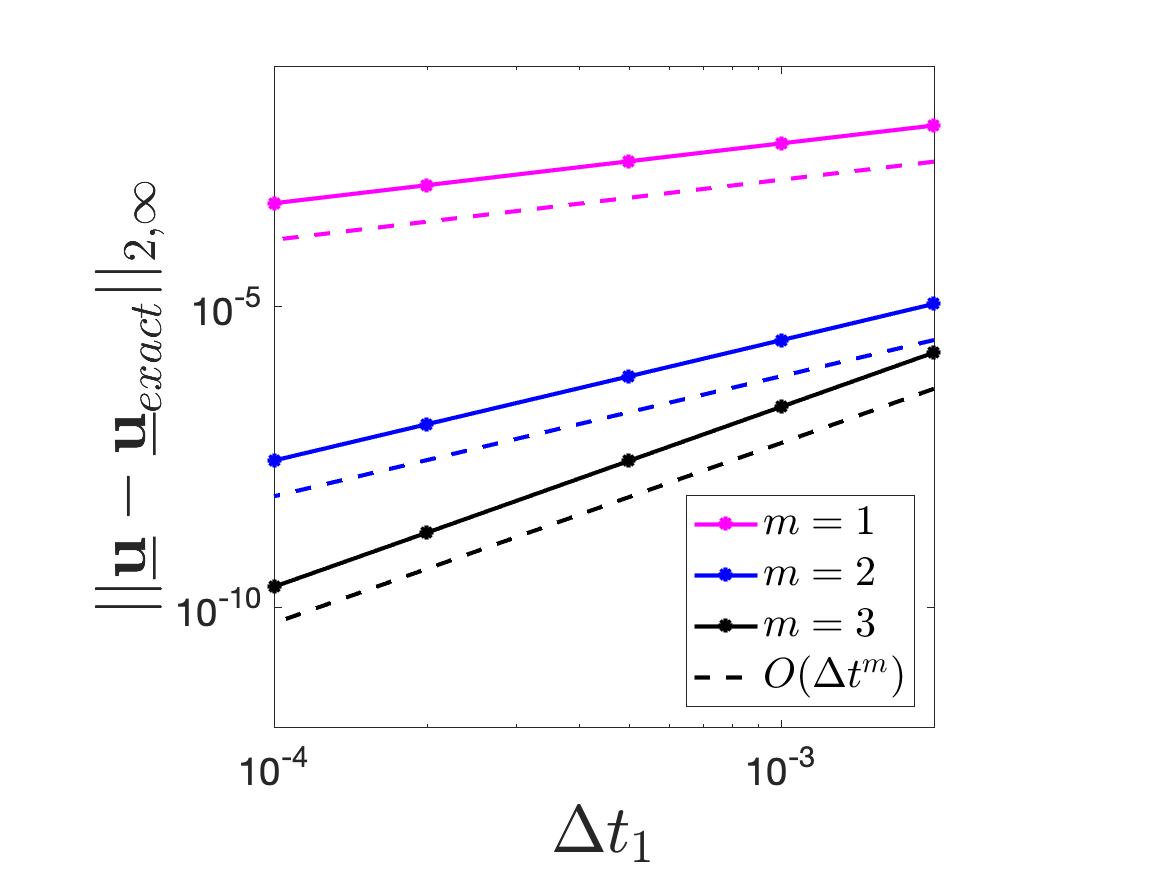} & \hspace{-5mm}
\includegraphics[height=45mm]{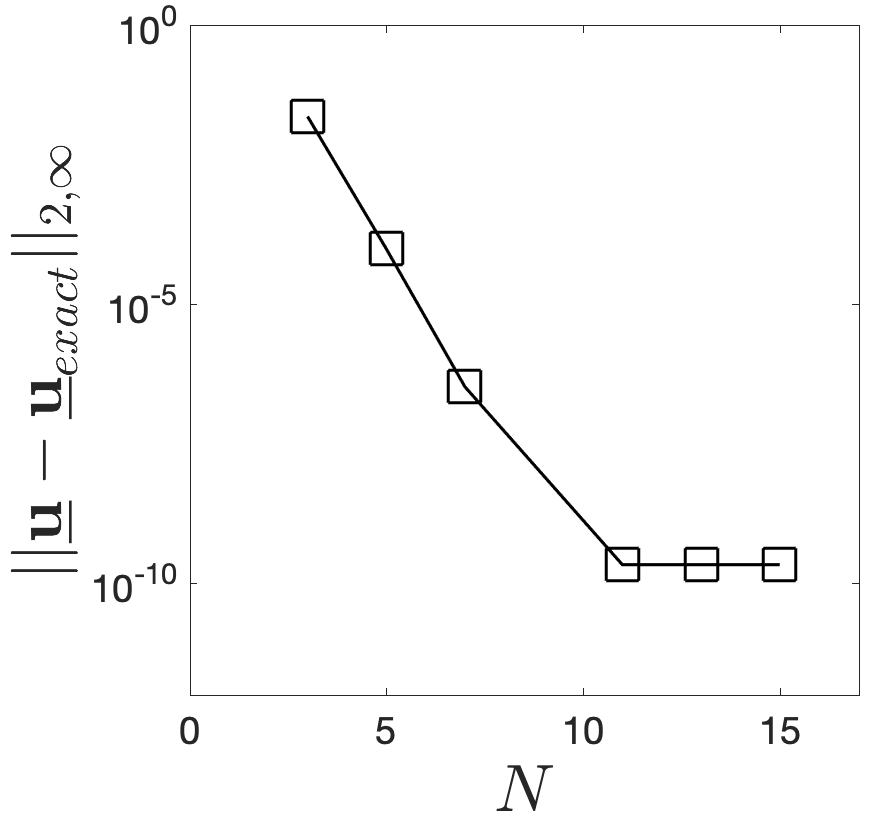} \\ \vspace{-2mm}
\textrm{(c) } & \hspace{-5mm} \textrm{(d) }
\end{array}$
\end{center}
\vspace{-7mm}
\caption{(a) Overlapping and (b) individual spectral element meshes for the doubly-periodic domain,
(c) temporal and (d) spatial convergence of the error in the solution.}
\label{fig:eddymeshes} \end{figure}

Here, we model a periodic domain $\Omega:=[0,2\pi]^2$ using three overlapping meshes that are illustrated Fig. \ref{fig:eddymeshes}(a). A doubly-periodic background mesh ($\Omega^1$ with $E=240$) has a square hole in the center that is covered with a pair of circular meshes ($\Omega^2$ with $E=96$ and $\Omega^3$ with $E=140$). The individual meshes are shown along with their interdomain boundaries in Fig. \ref{fig:eddymeshes}(b). The flow parameters are $\nu=0.05$, $\bu_0 = (1,0.3)$, ${\psi} = (1/5)sin(5y) +(1/5)cos(5x) - (1/4)sin(3x)sin(4y)$, and $\lambda=-25$.

To demonstrate temporal convergence of the MTS-based method, the flow is
integrated up to time $T_{f}=1$ convective time units (CTU) at different $\dt$
for extrapolation order $m=1$, 2, and $3$. To ensure stability of the predictor-corrector
approach for multirate timestepping, we set $Q=0$ for $m=1$, $Q=1$ for $m=2$, and
$Q=3$ for $m=3$. The interpolation order during the
correction iterations is set to $\tilde{m}={\tt max}(1,m-1)$. Additionally, because of the
difference in the size of the elements in the three meshes, the timestep ratio
is set to $\dt_1/\dt_2=2$ and $\dt_1/\dt_3=3$ to keep the CFL similar for the
three subdomains.
The polynomial order is $N=13$ for this convergence study and the
BDF3/EXT3 scheme is used for all the results presented here.

Figure \ref{fig:eddymeshes}(c) shows that the MTS-based method maintains the
temporal convergence of the underlying SEM solver. Here, the error is computed as
$\bue=\buu-\buu_{exact}$, and the norm is the 2-norm of the point-wise maximum
of the vector field, i.e., $||{\bf \ue}||_{2,\infty} :=
||{\tilde{\bold{e}}}||_2$, where
$\tilde{\bold{e}}=[||\underline{e}_1||_{\infty},||\underline{e}_2||_{\infty}]$.
For each $m=1$, 2 and 3, we observe that the error between the numerical solution
and the exact solution decreases as $O(\dt^m)$.

Similarly, for spatial convergence,
the flow is
integrated up to time $T_{f}=1$ convective time units (CTU) at different $N$.
The timestep size is fixed to $\dt_1=10^{-4}$ and
$Q=3$ corrector iterations are used
at each timestep with third-order extrapolation ($m=3$) for the interdomain boundary
data. Figure \ref{fig:eddymeshes}(d) shows the exponential convergence of the
solution obtained using MTS method with change in $N$.
The temporal and spatial convergence results presented here demonstrate that the
MTS method presented in this paper maintains the convergence properties of the underlying SEM
solver.

\subsection{Buoyant Thermal Plume} \label{sec:plume}

Buoyant plumes arise in a variety of industrial and environmental flow problems
such as deepwater blowouts \cite{reddy2012composition}, volcanoes \cite{kaminski2005turbulent} and hydrothermal vents \cite{walter2010rapid},
and they have been the subject of several experimental and computational
studies (e.g., \cite{
fabregat2016dynamics,schmidt1941turbulent,lavelle1997buoyancy,tomas2016effects,tomas2017numerical}).  As noted in Section \ref{sec:intro}, plumes often
feature significant scale disparity;  the high speed, highly turbulent, flow
near the plume or jet exit requires fine scale meshes and correspondingly small
timestep sizes in that region, whereas the far-field flow is typically
relatively quiescent, with larger-scale and slower turbulent eddies.  Thus,
buoyant plumes (and even non-buoyant jets) are ideal candidates for
discretizations that are multi-resolution both in space and time.  This point
is emphasized by the example of Fig. 1, which shows a multidomain spatial
discretrization where the near-field CFL is about 100 times larger than the
far-field CFL when using an STS.

To explore the potential of our Schwarz-MTS coupling for this class of
problems, we consider a singlephase thermally-buoyant plume in a stratified
environment.
In this example, we assume that there is a reference density ($\rho_r$) and a reference temperature
($T_r$), with respect to which the density varies in the domain as
$\rho=\rho_r(1-\gamma(T-T_r))$, where $\gamma$ is the thermal expansion
coefficient of the fluid. We also assume that the temperature in the domain,
which is a solution of (\ref{eq:nseenergy}), can be described as
$T(\bx,t)={\theta}(\bx,t)+T_r+\Gamma z$, where $\theta$ is the perturbation with
respect to the unperturbed environment temperature, $T_e$, which varies linearly
with a slope of $\Gamma$ as $T_e = T_r + \Gamma z$. Here, $z$ is the direction
in which the fluid is stratified, which is always in opposite direction to the
gravitational acceleration for stable stratification.

The effect of variation in density due to the temperature difference in the fluid
and the transport of $\theta$ in the domain
is modeled using the Boussinesq approximation \cite{kundu2001fluid}. These assumptions
lead to a system of the form
\begin{eqnarray}
\label{eq:plmomentum}
 \qquad \frac{\partial \bu}{\partial t}  + \bu.\nabla{\bu} &=& - \nabla{p} +
\frac{1}{Re} \nabla^2{\bu} + Ri \theta \hat{k},  \\
\label{eq:plmass}
\nabla\cdot \bu &=& 0, \\
\label{eq:plenergy}
\frac{\partial \theta}{\partial t}  + \bu.\nabla{\theta} &=&
\frac{1}{Pe} \nabla^2{\theta} - \bu \cdot \hat{k},
\end{eqnarray}
which we solve using the Schwarz-SEM framework.
In (\ref{eq:plmomentum}), $Ri=g/(B_o^{1/4}N_b^{5/4})$ is the Richardson number
that depends on the acceleration due to gravity ($g$), inlet buoyancy flux ($B_0$),
and the buoyancy frequency ($N_b$) (also known as the Brunt–V\"{a}is\"{a}l\"{a} frequency).
The velocity, time, length, pressure, and temperature scales used for
nondimensionalization are $U_0=(B_0N_b)^{1/4}$, $t_0=1/N_b$, $L_0=(B_0/N_b^3)^{1/4}$,
$p_0=\rho_rU_0^2$, and $T_0=\Gamma L_0=(B_0N_b^5)^{1/4}/(g\gamma)$, where $g$
is the acceleration due to gravity.  We assume that the reference density is
$\rho_r=1$ and the reference temperature is $T_r=0$. The temperature solution
$T(\bx,t)$ can be obtained by substituting $T(\bx,t)={\theta}(\bx,t)+T_r+\Gamma z$.

Here, we follow \cite{fabregat2016dynamics}
and specify $B_0=5\times10^{-6} m^4s^{-3}$ , $N_b=0.1s^{-1}$.
The linear scaling for density is set to $\gamma=2\times10^{-4} K^{-1}$, and
for temperature to $\Gamma=5.1m^{-1}$.  Double diffusion effects are ignored,
and thus $\nu=\alpha=10^{-6}m^2s^{-1}$, which leads to Prandtl number $Pr=1$.
The Reynolds number ($U_0L_0/\nu$) and Peclet number of the flow is about 7100,
and the Richardson number is 3700. The reader is referred to Fabregat et al. \cite{
fabregat2016dynamics} for a detailed derivation of the governing equations for
this example.

\begin{figure}[t!] \begin{center}
$\begin{array}{c}
\includegraphics[height=80mm]{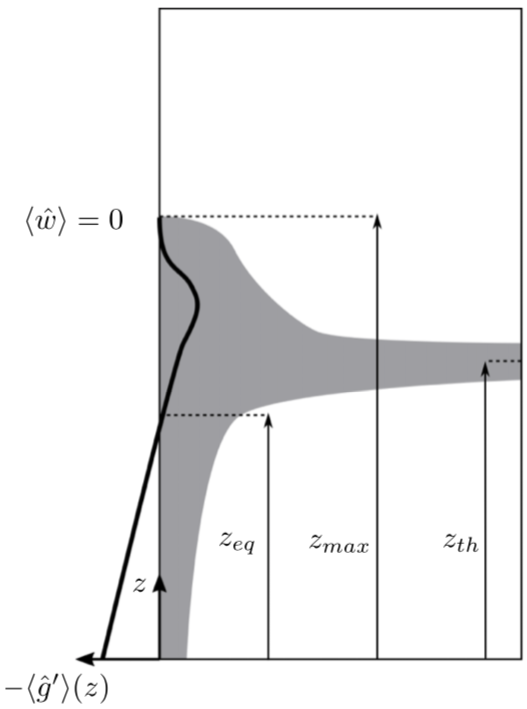}
\end{array}$
\end{center}
\caption{Schematic of a singlephase plume indicating the maximum plume height ($z_{max}$),
 trapping height ($z_{th}$) and equilibrium height($z_{eq}$). Image
taken from \cite{fabregat2016dynamics}.}
\label{fig:plumeschematic}
\end{figure}

To validate the MTS scheme, we are interested in accurately determining
three key plume parameters (illustrated in Fig.
\ref{fig:plumeschematic}) and comparing
these results from MTS with the monodomain SEM framework and \cite{fabregat2016dynamics}:
\begin{itemize}
\item Maximum height $z_{max}$ - the maximum height at which the axial velocity
of the plume vanishes.
\item Trapping height $z_{th}$ - the height of
the centerline of the outgoing gravity current.
\item Equilibrium height $z_{eq}$ - the height at which the plume becomes
neutrally buoyant.
\end{itemize}

\noindent
Note that the results presented in \cite{fabregat2016dynamics} were also obtained
using the monodomain SEM framework. The only difference in the problem setup of
\cite{fabregat2016dynamics} from the ongoing calculations is that
Fabregat et al. do not model the recycling-pipe attached at the
bottom of the cylindrical tank (Fig. \ref{fig:mono_plume}), and instead applied inhomogeneous Dirichlet
condition (defined an inlet velocity) directly at the bottom of the tank. The
recycling-pipe inlet was implemented in the current setup to
allow for fully developed turbulent inflow
in the plume. Though, difference in the inlet setup between
\cite{fabregat2016dynamics} and the current study will not have appreciable
affect on the three parameters defined above, as every correctly simulated stable
turbulent buoyant plume converges to the asymptotic solution derived by
Morton-Taylor-Turner \cite{morton1956}.

\subsubsection{Results}
Figure \ref{fig:mono_plume}(b,c)-(d) shows the spectral element meshes that were used for the monodomain and Schwarz-SEM calculations, respectively. The conforming mesh for monodomain SEM has 76,600 elements, and the overlapping spectral element meshes have a total of 71,040 elements. $E_f=55,480$ elements for the dense inner grid ($\Omega^f$) and $E_c=15,560$ elements in the coarse outer grid ($\Omega^c$). The total element count is lower for the Schwarz-SEM framework because the overlapping meshes are nonconforming with the outer mesh much coarser as compared to the inner-mesh.

Using the multirate timestepping method described in Section \ref{sec:method},
two different timestep ratio are used for the Schwarz-SEM framework; $\tsr=5$
and $\tsr=100$. Since the subdomain with slower time-scales has to take many
fewer timesteps in the MTS-based scheme, we use fewer MPI ranks for the outer
domain in comparison to the MPI ranks needed for the STS-based scheme. The
timing analysis for the MTS-based scheme has been presented in Section
\ref{sec:timinigmulti}.

\begin{figure}[b!]
\begin{center}
$\begin{array}{cccc}
\includegraphics[width=12mm]{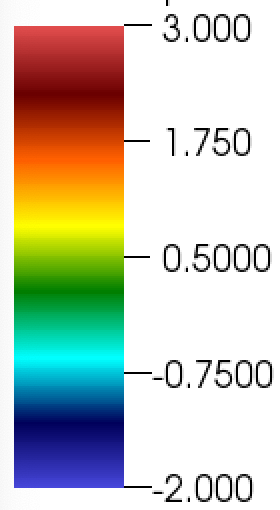} &
\includegraphics[width=55mm]{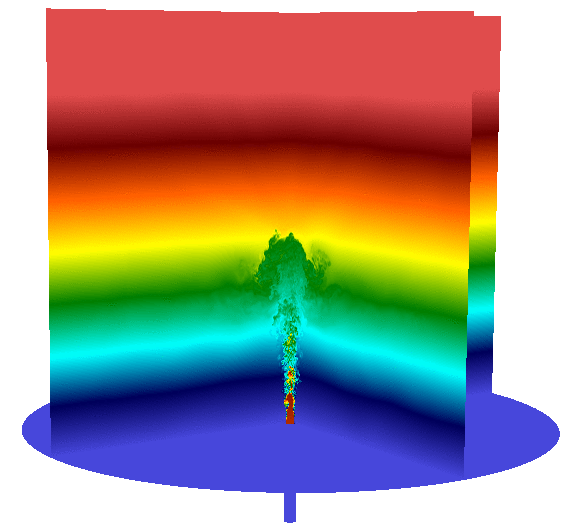} &
\includegraphics[width=55mm]{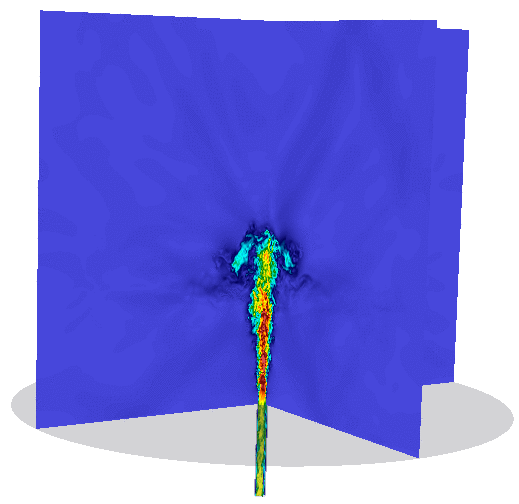} &
\includegraphics[width=12mm]{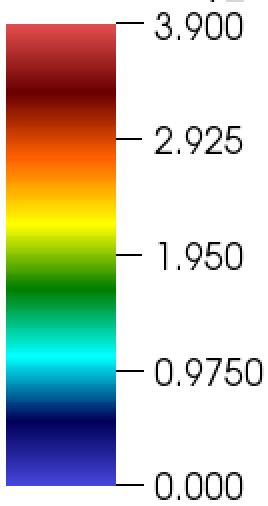}  \\
\end{array}$
\end{center}
\vspace{-5mm}
\caption{Snapshot of instantaneous (a) temperature and (b) velocity magnitude
solution from the overlapping grid calculation with $\tsr=100$.}
\label{fig:plumeresults3d}
\end{figure}
Following the monodomain calculation, the polynomial order is set to $N=7$ for the overlapping grid calculation. Since the overlap region is away from the area of interest, we set $Q=0$ with $m=1$, and the flow statistics are temporally-averaged over more than 30 CTU.
Figure \ref{fig:plumeresults3d} shows a snapshot of the instantaneous temperature
and velocity magnitude solution from the overlapping grid calculation with $\tsr=100$.
\begin{figure} \begin{center}
$\begin{array}{cccc}
\includegraphics[width=35mm]{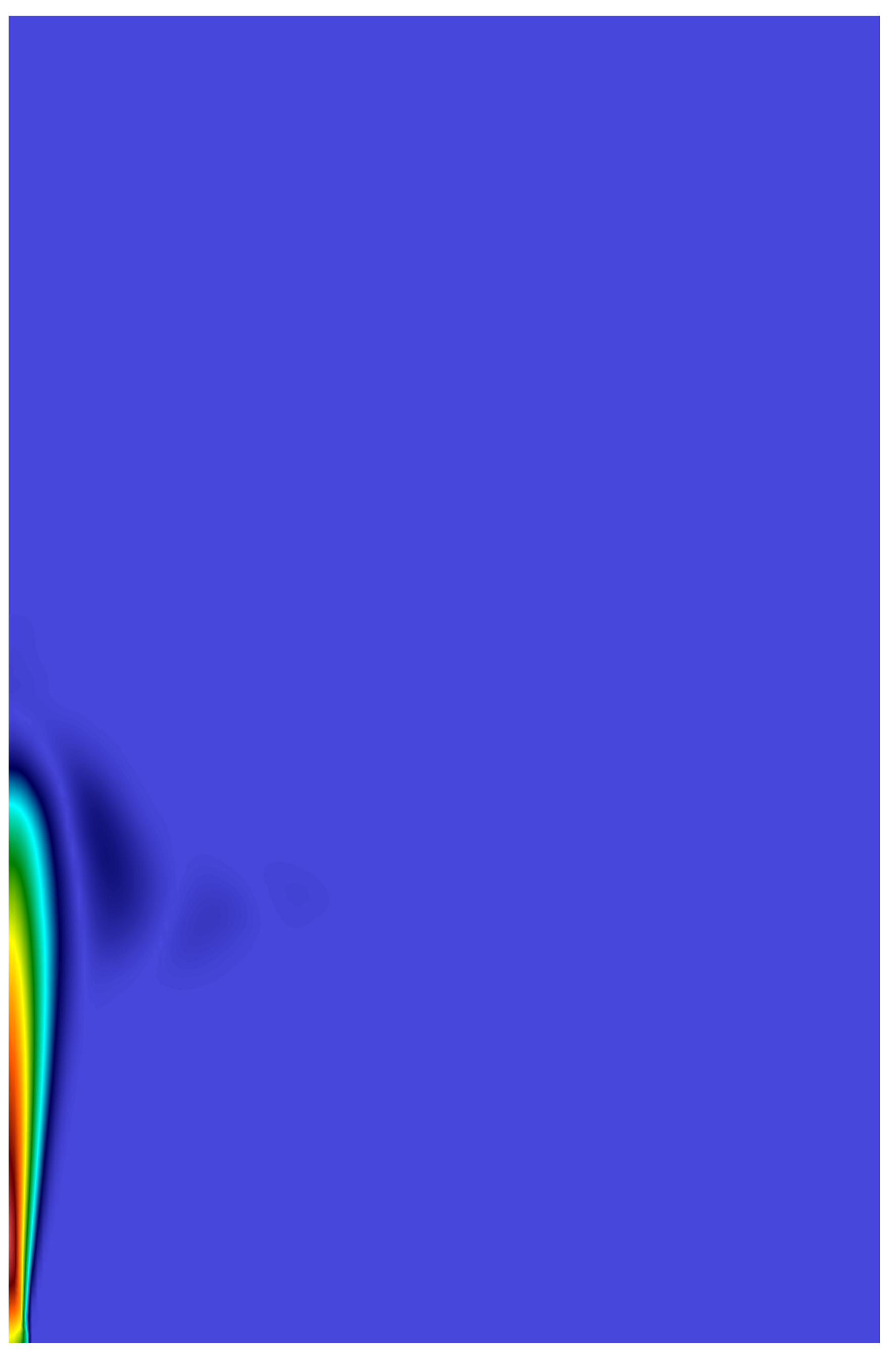} &
\includegraphics[width=35mm]{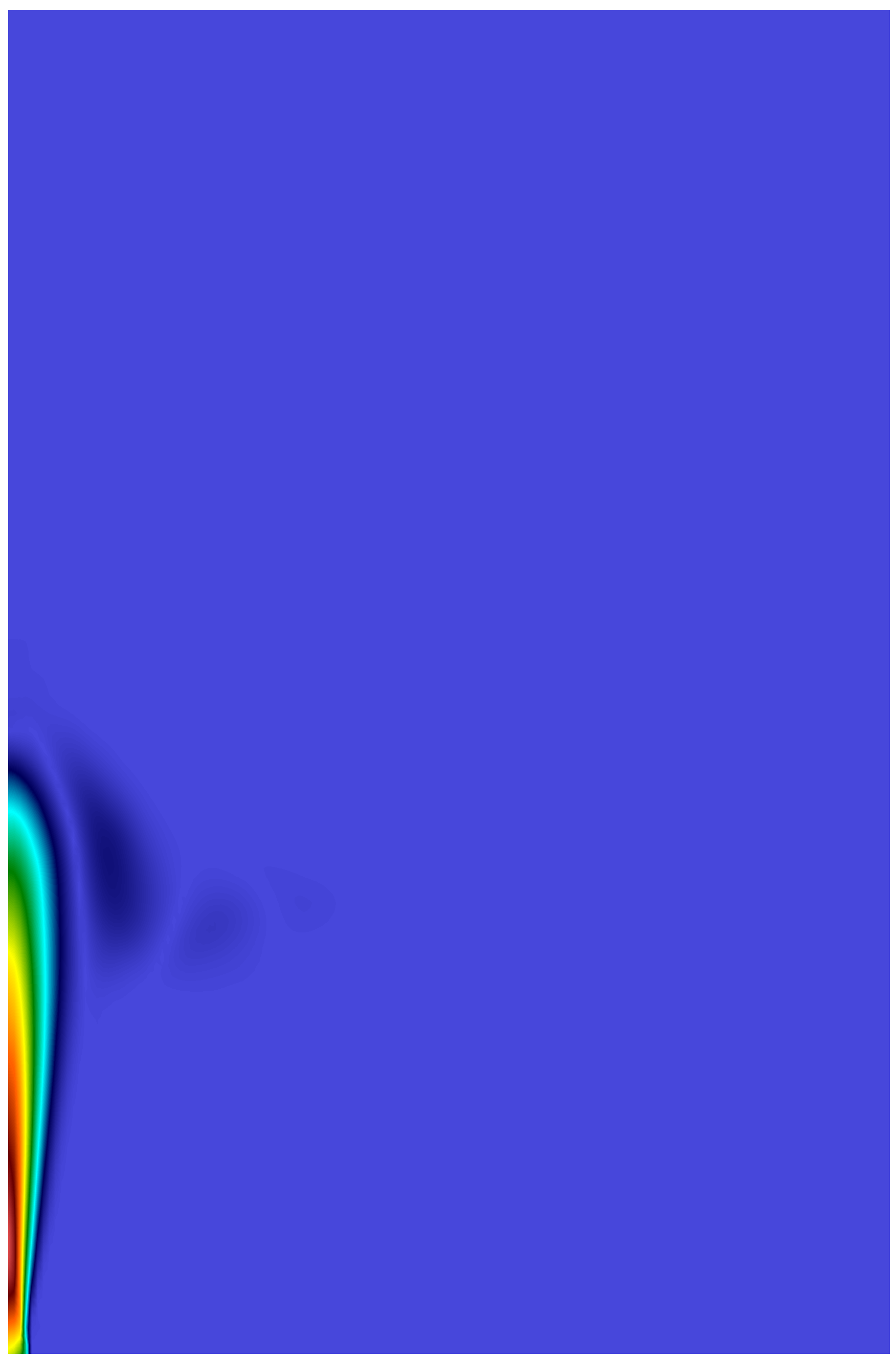} &
\includegraphics[width=35mm]{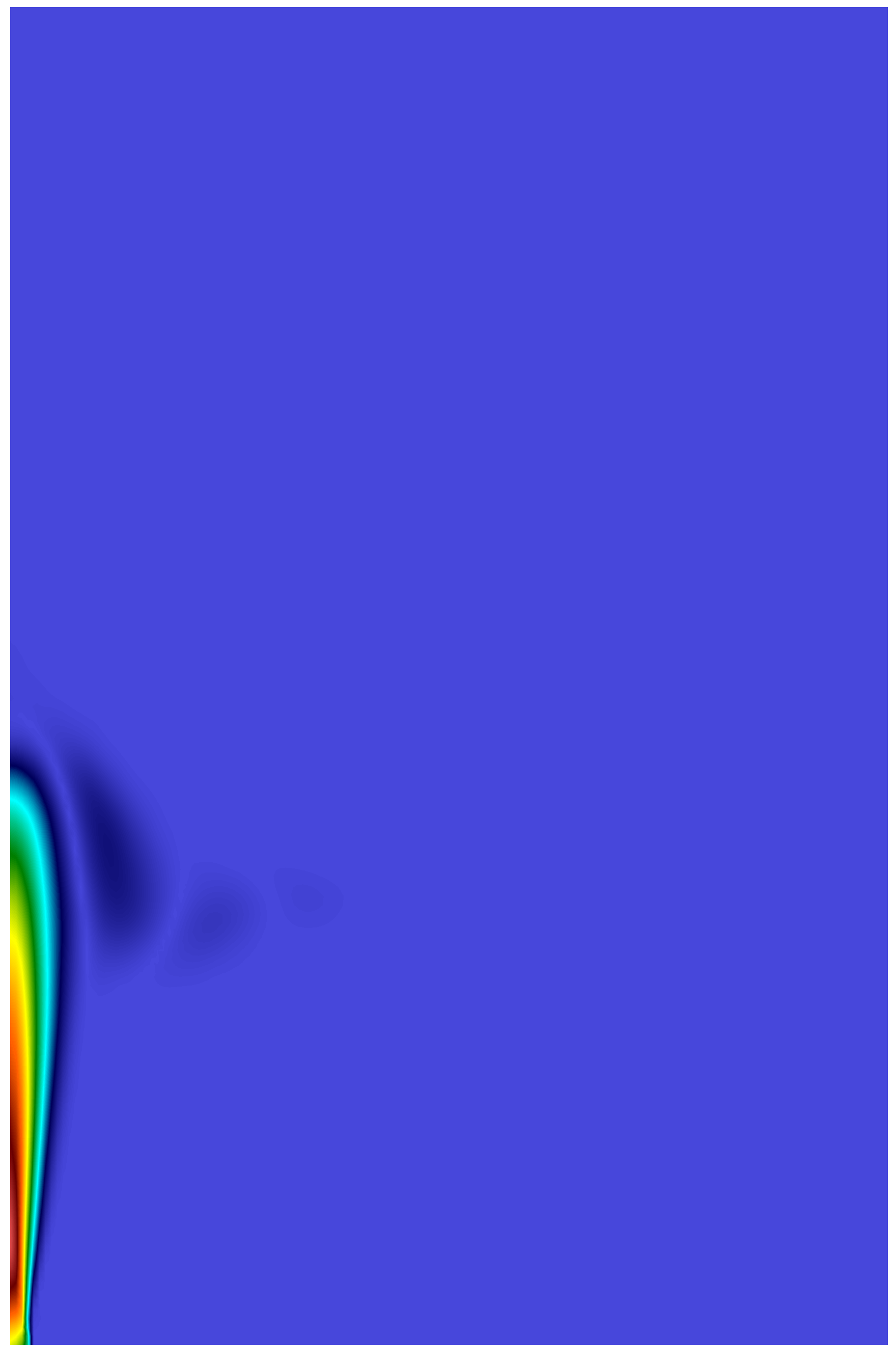} &
\includegraphics[width=20mm]{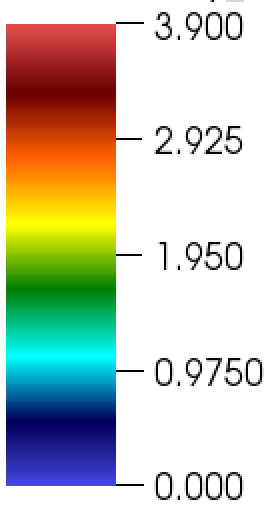}  \\
\multicolumn{4}{c}{\textrm{(a) Velocity magnitude contours}}  \\
\includegraphics[width=35mm]{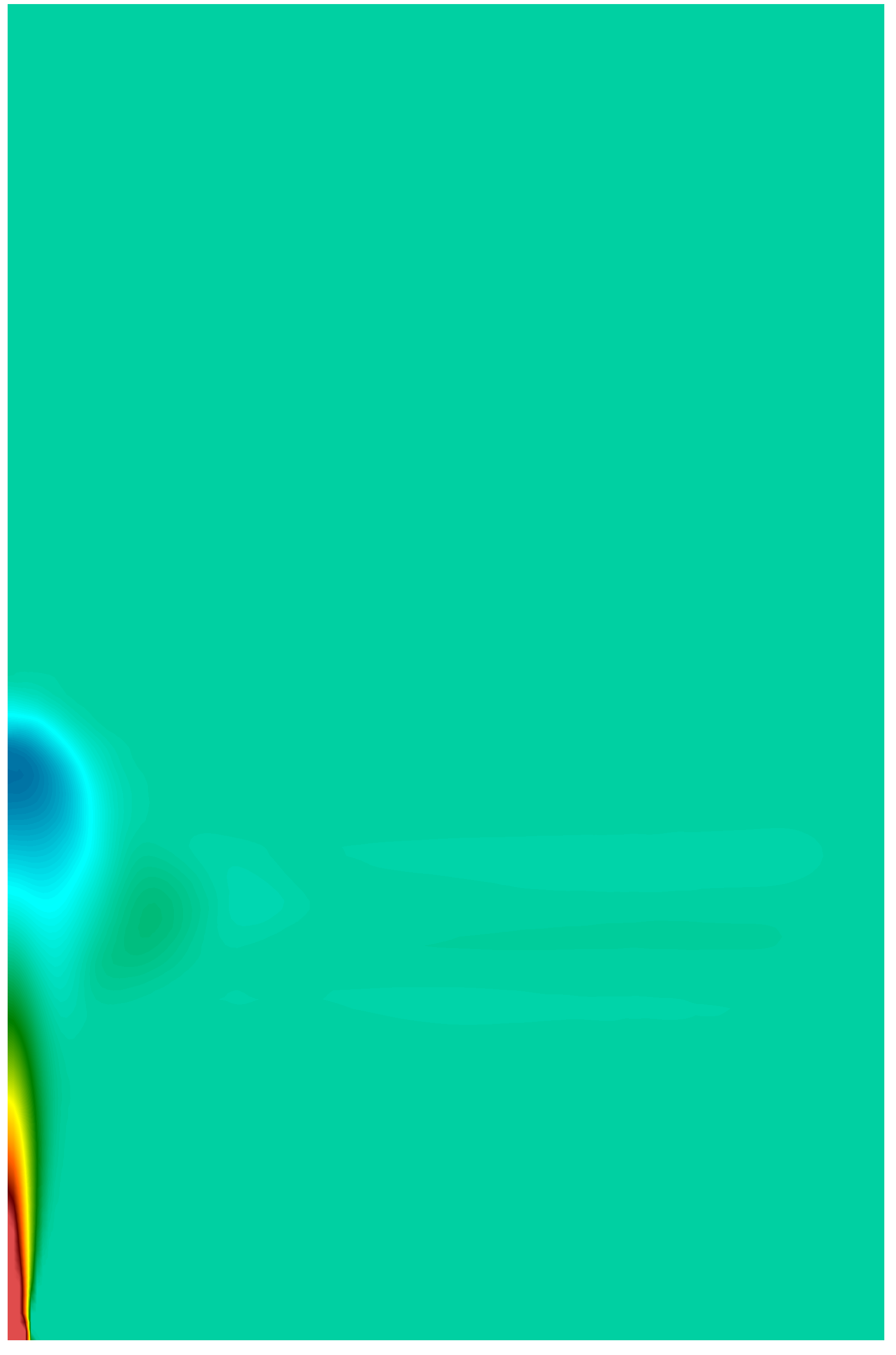} &
\includegraphics[width=35mm]{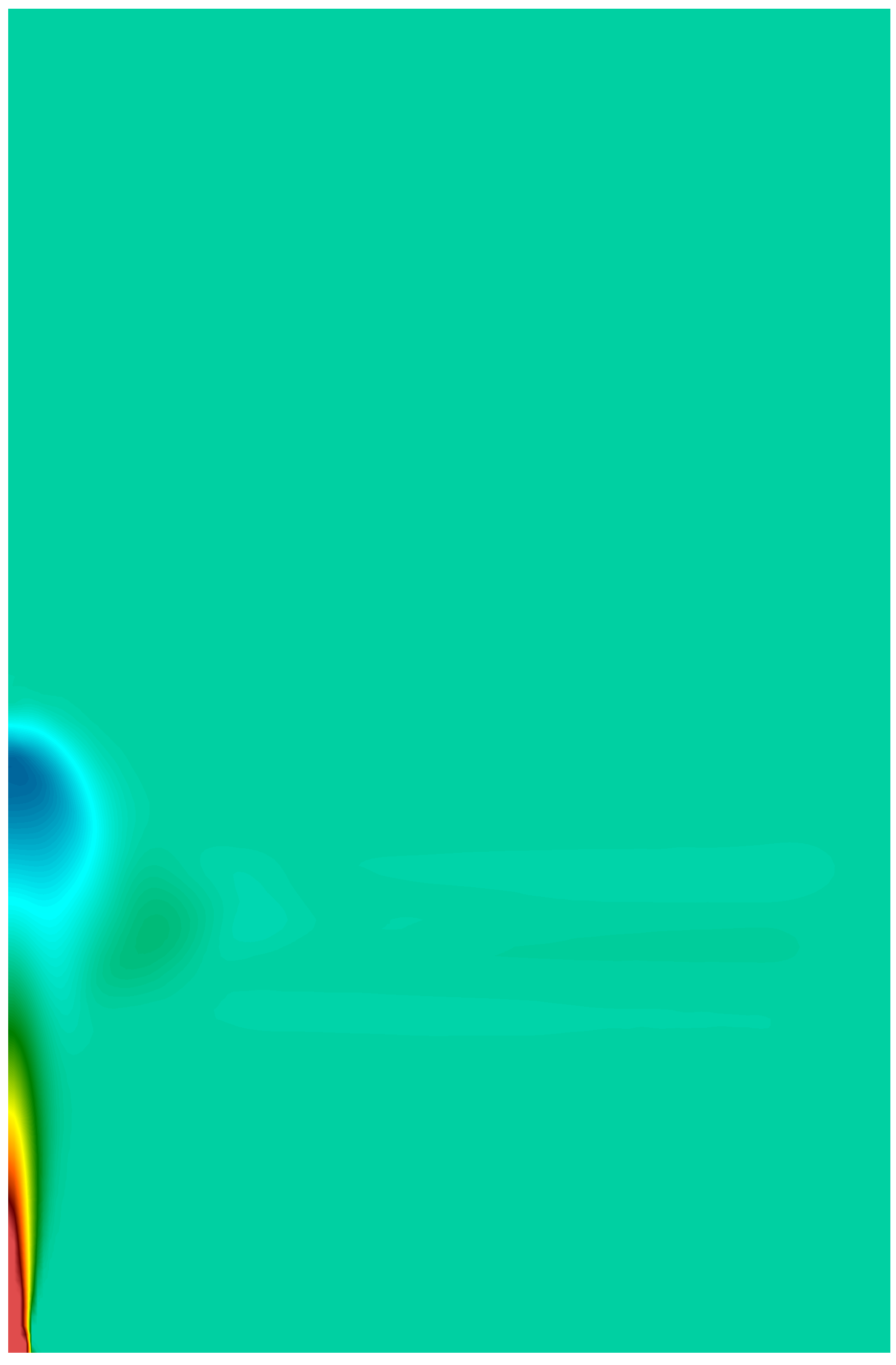} &
\includegraphics[width=35mm]{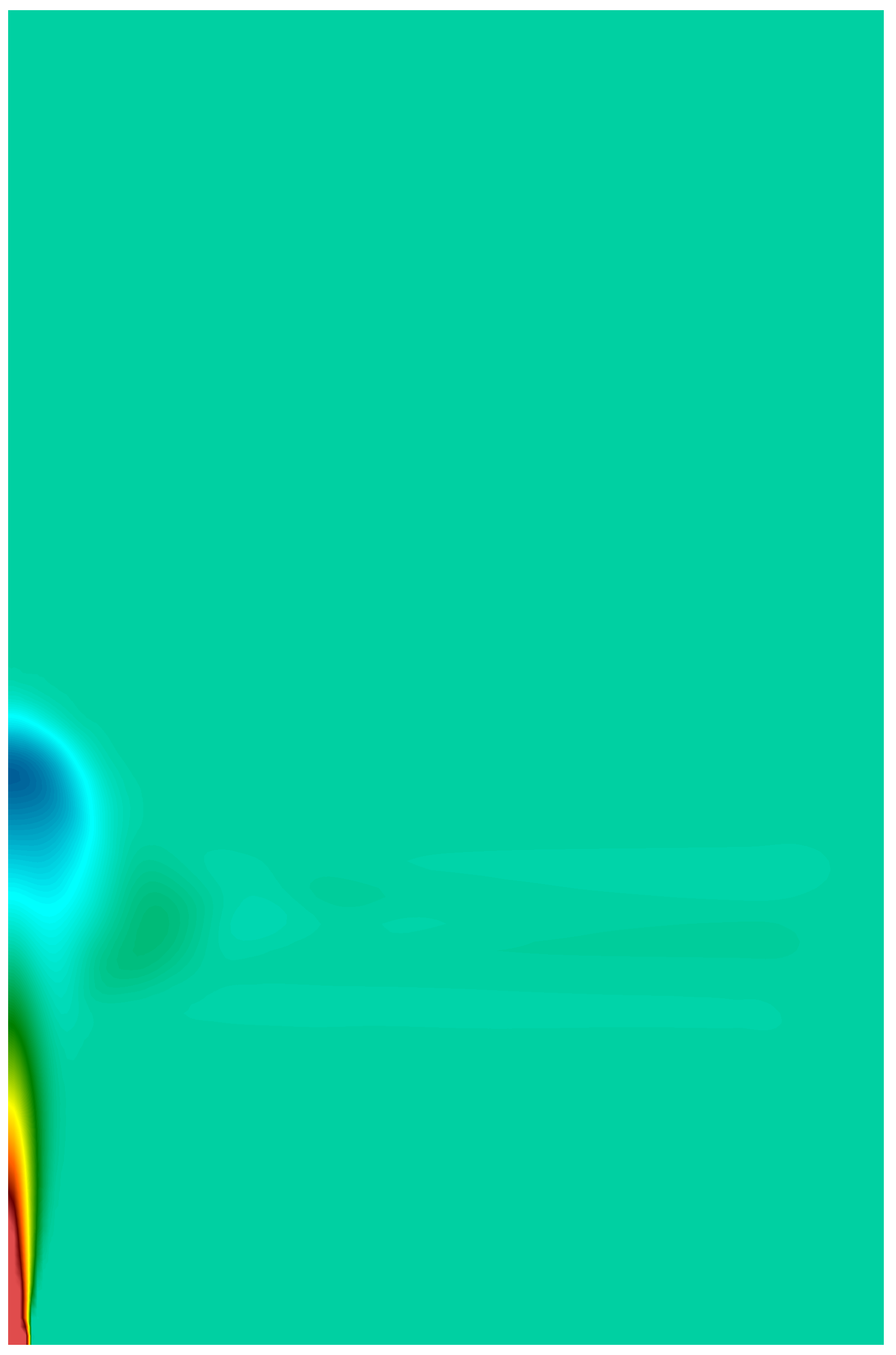} &
\includegraphics[width=20mm]{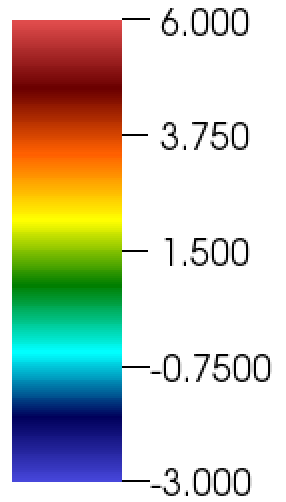} \\
\multicolumn{4}{c}{\textrm{(b) Temperature perturbation contours}}  \\
\includegraphics[width=35mm]{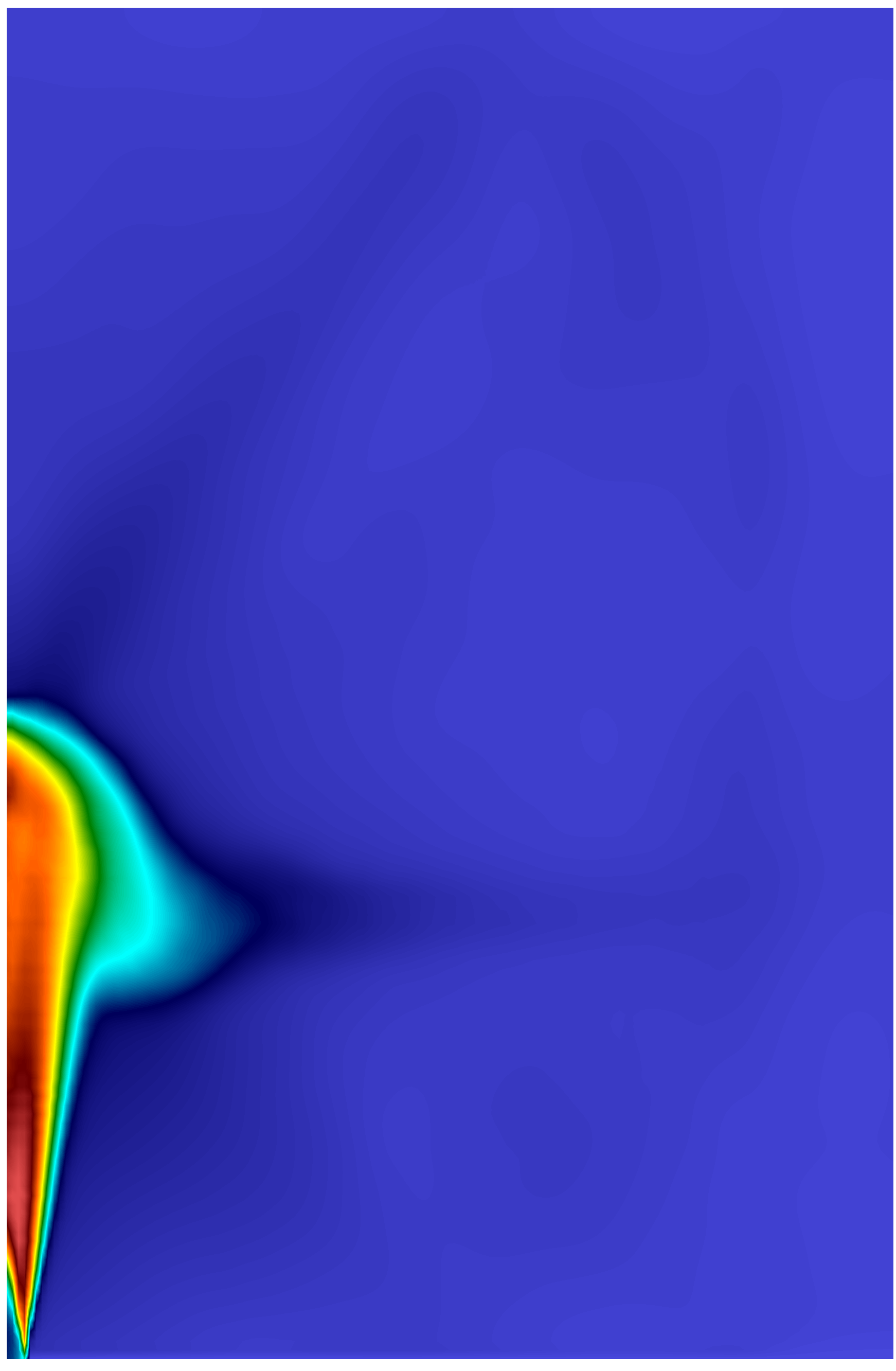} &
\includegraphics[width=35mm]{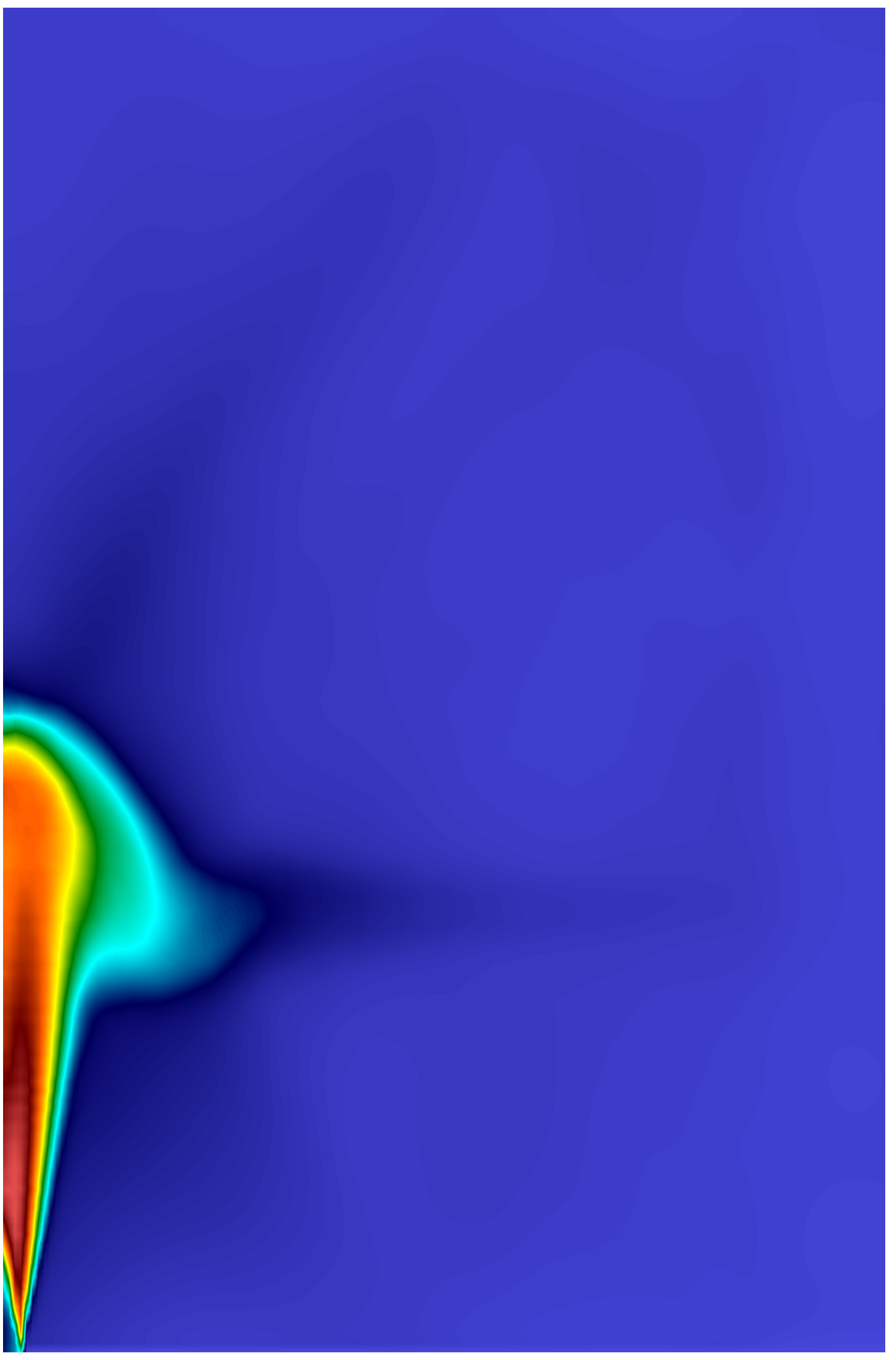} &
\includegraphics[width=35mm]{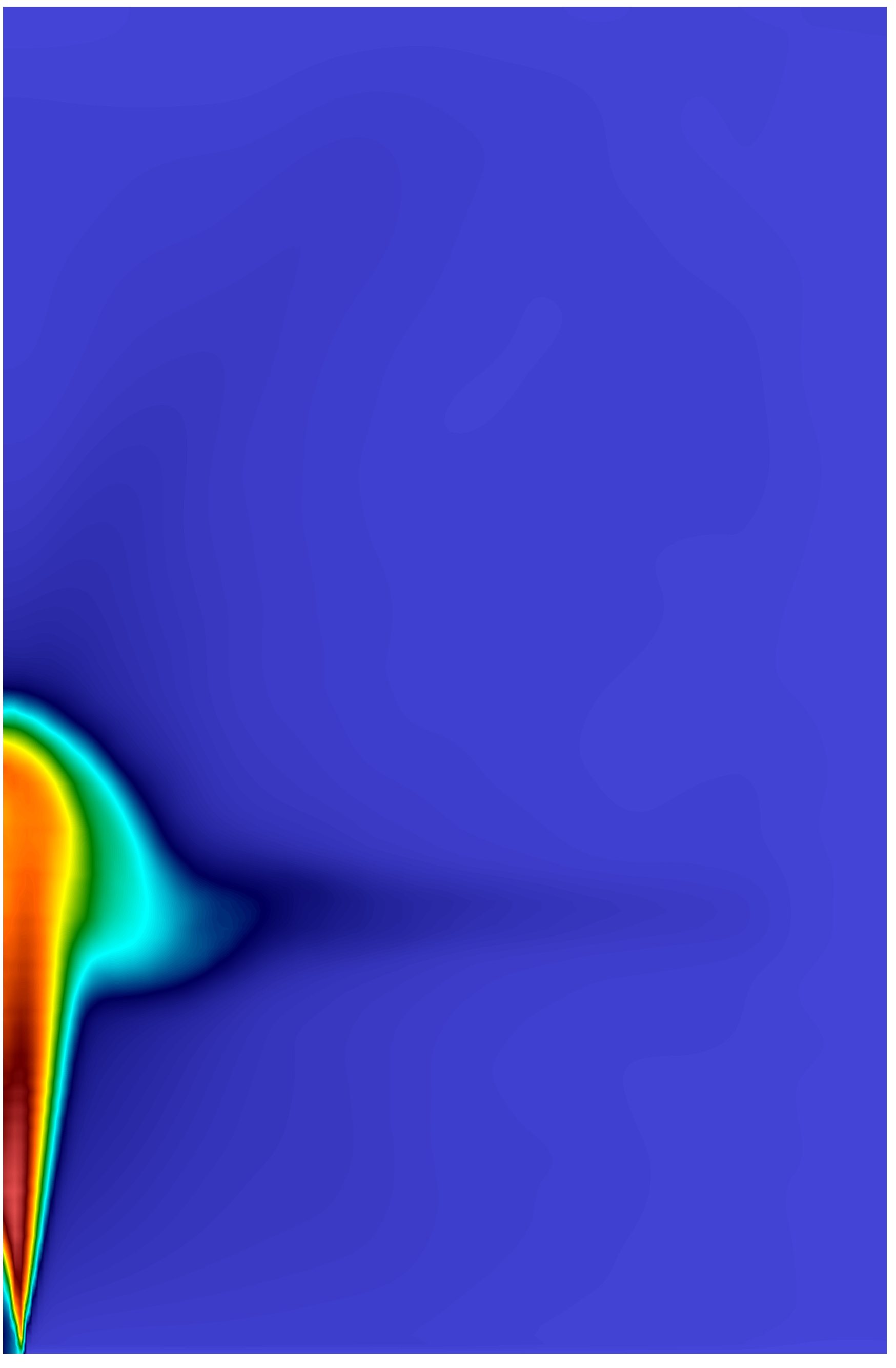} &
\includegraphics[width=20mm]{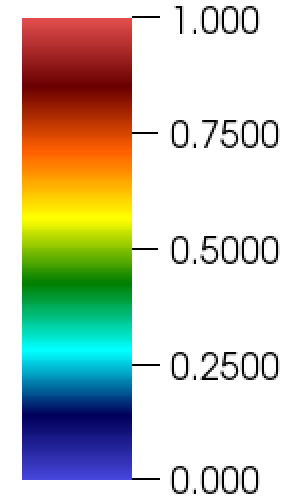} \\
\multicolumn{4}{c}{\textrm{(c) Turbulent kinetic energy contours}}
\end{array}$
\end{center}
\vspace{-5mm}
\caption{Temporally and spatially averaged (top to bottom) (a) velocity
magnitude, (b) temperature perturbation, and (c)
turbulent kinetic energy for the Schwarz-SEM calculations with overlapping
grids using (left) $\tsr=5$ and (center)
$\tsr=100$, and (right) the monodomain SEM calculation.}
\label{fig:plumeresults}
\end{figure}
Figure \ref{fig:plumeresults} shows the temporally and spatially averaged (azimuthally averaged) velocity magnitude, temperature perturbation, and turbulent kinetic energy (TKE) contours, respectively, for the overlapping grid ($\tsr=5$ and 100) and monodomain calculations. We observe that the Schwarz-SEM framework gives good comparison with the monodomain SEM calculation even when the timestep ratio is 100.

\begin{figure}[t!] \begin{center}
$\begin{array}{cc}
\includegraphics[width=60mm]{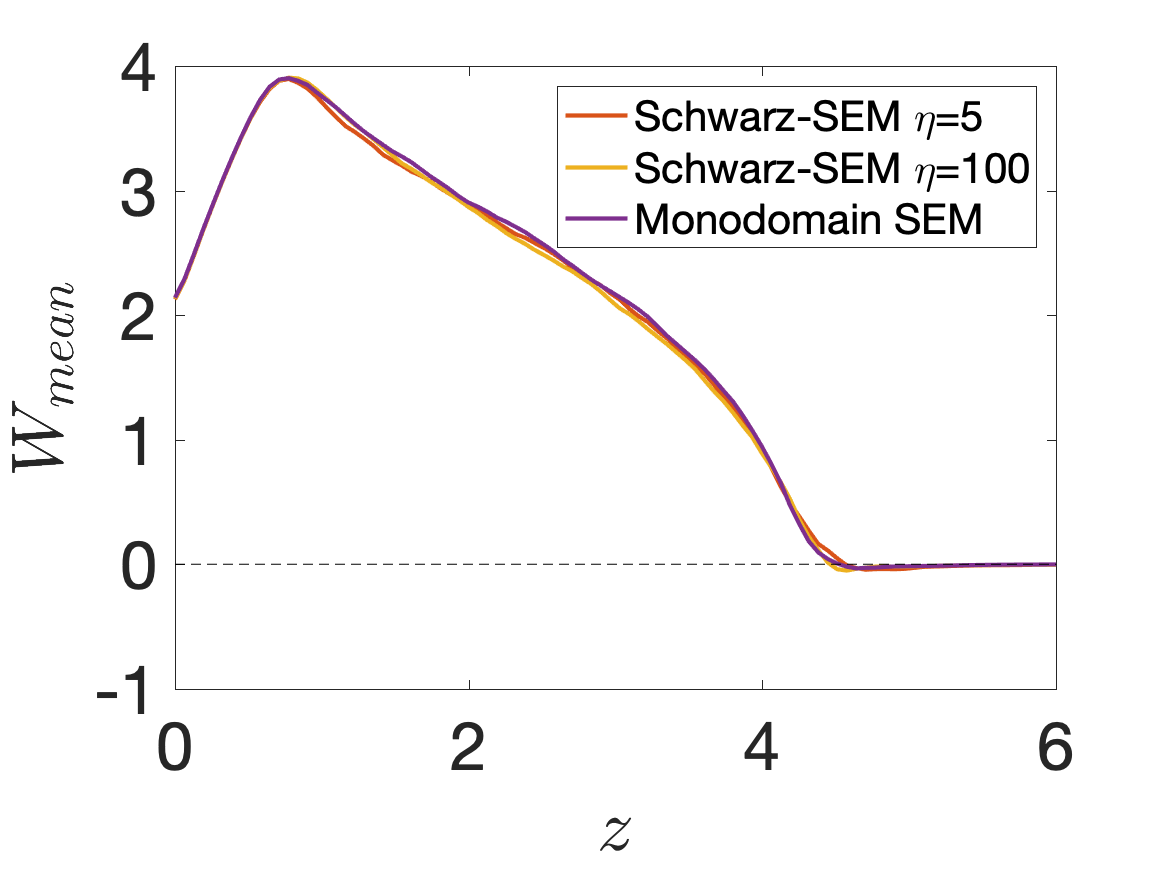} &
\includegraphics[width=60mm]{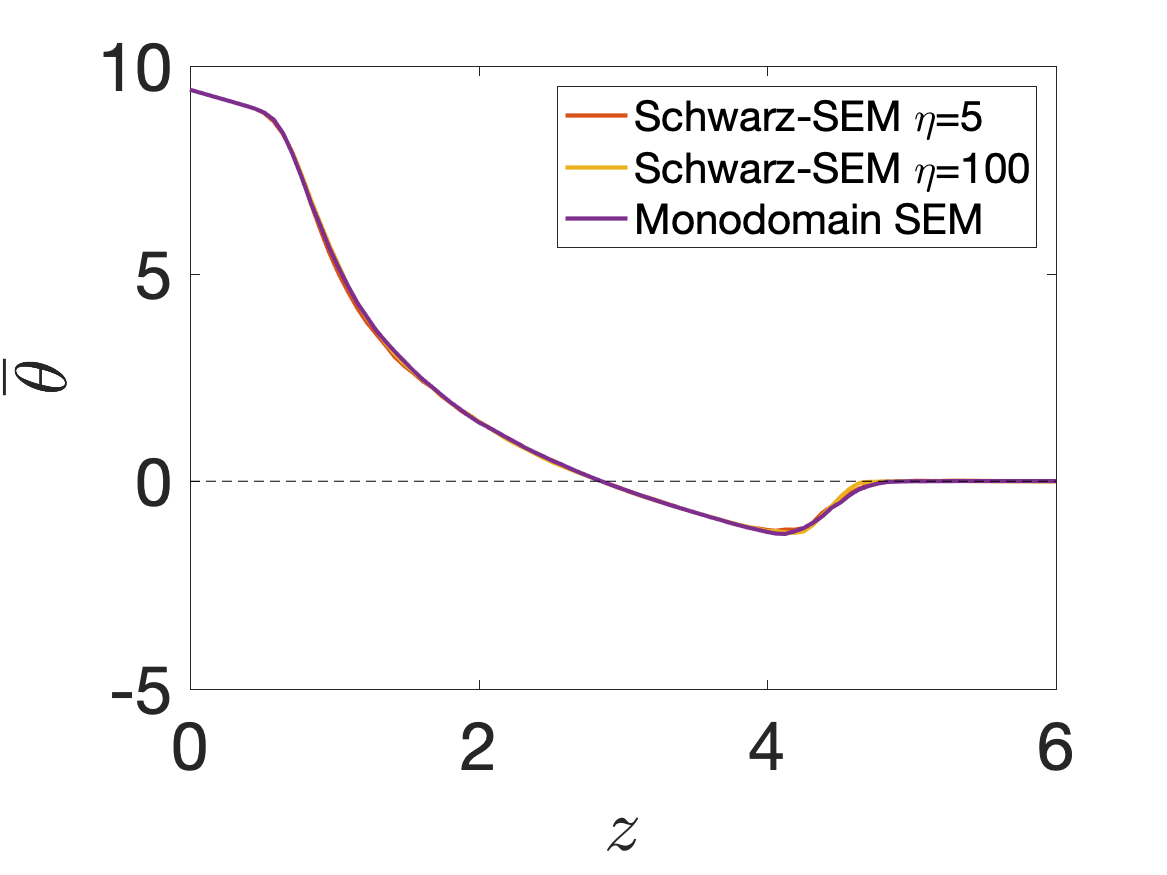} \\
\end{array}$
\end{center}
\caption{Temporally and spatially averaged (left) axial velocity ($W_{mean}$)
and
(right) temperature perturbation ($\theta_{mean}$) along the plume centerline.}
\label{fig:plumevtcenter}
\end{figure}

Using the temporally and spatially averaged axial velocity ($W_{mean}$)
and temperature perturbation (${\theta_{mean}}$) along the plume centerline, we can obtain
the maximum height of the plume ($z_{max}$) and equilibrium height ($z_{eq}$). We can also use the
TKE plots to obtain the plume trapping height ($z_{th}$). The line plots
comparing $W_{mean}$ and ${\theta_{mean}}$ are shown in Fig.  \ref{fig:plumevtcenter},
and show good comparison between the
three cases considered here.

\begin{table}[t!] \begin{center}
 \begin{tabular}{|r|c|c|c|}
 \hline
  & $z_{max}$ & $z_{eq}$ & $z_{th}$ \\
 \hline\hline
 Fabregat et al. & 4.5 & 2.85 & 3.11 \\
 \hline
 Monodomain SEM & 4.51 & 2.85 & 3.15 \\
 \hline
 Schwarz-SEM $\tsr=5$ & 4.56 & 2.85 & 3.15 \\
 \hline
 Schwarz-SEM $\tsr=100$ & 4.46 & 2.85 & 3.19 \\
 \hline
\end{tabular}
\caption{$z_{max}$, $z_{eq}$ and
$z_{th}$ obtained from the Schwarz-SEM framework, the current monodomain
calculation, and the monodomain calculation by Fabregat \cite{fabregat2016dynamics}.}
\label{table:plumeheights}
\end{center}
\end{table}

Table \ref{table:plumeheights} compares the maximum plume height ($z_{max}$),
equilibrium height ($z_{eq}$), and trapping height ($z_{th}$) for the
Schwarz-SEM results with monodomain calculations and the monodomain SEM results
of Fabregat et al.  \cite{fabregat2016dynamics}. The maximum difference between
the Schwarz-SEM calculations and monodomain calculation for the three
parameters of interest is 1.2\% (for $z_{max}$). The
trapping height ($z_{th}$) for our monodomain calculation is
different from Fabregat et al., and that is expected as the flow was not fully
turbulent in the plume in \cite{fabregat2016dynamics}.

The results presented in this section demonstrate the effectiveness of the MTS
method in simulating complex turbulent flow and heat transfer phenomena. This MTS
method will be used for understanding the behavior of singlephase and multiphase
rotating plumes in a cross-flow. This target problem is intractable with a monodomain SEM framework
because a conforming mesh leads to a high element count, an issue that overlapping
grids help us circumvent.

\subsection{Timing Comparison between STS and MTS Method} \label{sec:timinigmulti}
With multirate timestepping, the subdomain with faster time-scales
($\Omega^f$) uses a smaller timestep size with more timesteps in comparison to the subdomain with
slower time-scales ($\Omega^c$).

Here, we use the thermally-buoyant problem with $\tsr=5$ to demonstrate
that multirate timestepping reduces the computational cost in comparison to
the corresponding singlerate timestepping-based ($\tsr=1$) calculation. For this example,
$E_f=55,480$ elements for the dense inner grid ($\Omega^f$) and $E_c=15,560$
elements in the coarse outer grid ($\Omega^c$).  For overlapping subdomains,
ideally one would partition the domain in parallel such that the time to
solution per timestep ($T_{step}$) is similar for each subdomain. For the
singlerate timestepping scheme, we typically choose the number of
MPI ranks ($P$) for each subdomain using
\begin{eqnarray}
\label{eq:singlerateproc}
\frac{P_c}{P_f} \approx \frac{E_c}{E_f},
\end{eqnarray}
where $P_f$ and $P_c$ are the number of MPI ranks use to partition $\Omega^f$
and $\Omega^c$, respectively. Based on $E_c$ and $E_f$ for this example, we
set $P_c \approx P_f/4$.  For the multirate scheme, however, since $\Omega^c$
has many times fewer steps as compared to the $\Omega^f$, the number of MPI
ranks for $\Omega^c$ can be reduced even further.

\begin{figure}[t!]
\begin{center}
$\begin{array}{c}
\includegraphics[height=60mm]{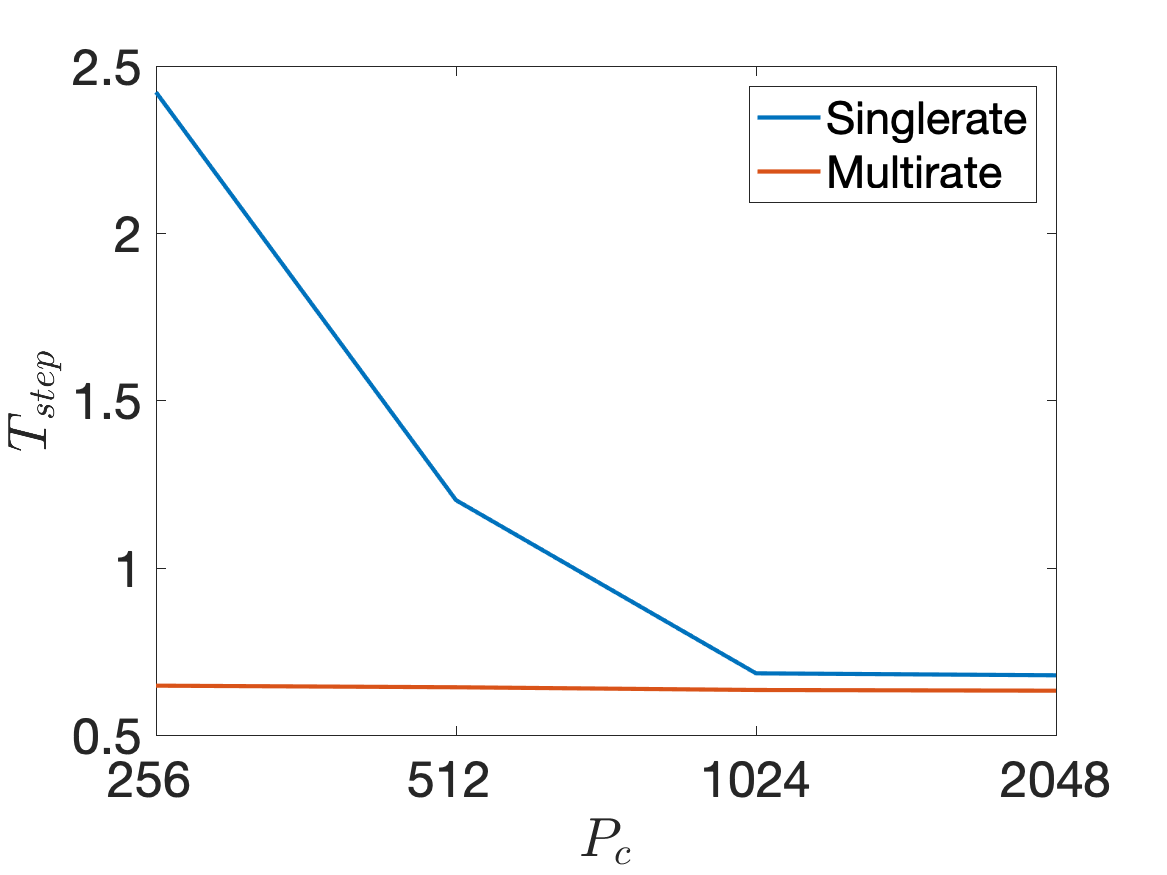}
\end{array}$
\end{center}
\caption{Variation of time to solution per timestep ($T_{step}$) for the
singlerate and multirate timestepping scheme with number of MPI ranks ($P_c$)
used for the subdomain ($\Omega^c$) with slower time-scales.  These tests were
done using the buoyant plume problem (Section \ref{sec:plume}), where
$E_f=55,480$ and $E_c=15,560$, and $P_f$ is fixed at $4096$ MPI ranks.}
\label{fig:plumemultitime}
\end{figure}

Figure \ref{fig:plumemultitime} compares how the mean time to solution per timestep
($T_{step}$) varies with $P_c$ for
the singlerate and multirate timestepping scheme, while keeping $P_f$ fixed at
4096 MPI ranks.  These calculations were done with $m=1$, $Q=0$, and $N=7$. The
time per timestep was obtained for multirate timestepping scheme by
monitoring the mean time taken by $\Omega^f$ for each sub-timestep, which is
equivalent to a single timestep in the singlerate timestepping scheme. The
timestep size was kept same for $\Omega^f$ for the multirate and singlerate
timestepping scheme, to ensure fair comparison. The numerical
experiments discussed here were done on Cetus, an IBM Blue Gene/Q at the
Argonne Leadership Computing Facility

The singlerate timestepping scheme is most efficient when
$P_c=1024=P_f/4$, which is in agreement with \eqref{eq:singlerateproc} since
$E_c\approx E_f/4$. As $P_c$ is decreased, the time to solution increases as
expected.  We also notice that increase in $P_c$ beyond 1024 does not change or
decrease the time to solution, which is due to $T_{step}$ being limited by
$\Omega^f$ once $P_c>P_f/4$.  In contrast to the singlerate scheme, since
$\Omega^c$ has to take fewer timesteps with the multirate timestepping
scheme, $P_c=P_f/16$ is as effective as $P_c=P_f/4$. $P_c$ cannot be reduced
further because of the constraint on maximum memory that can be allocated on
each MPI rank on Cetus.  Additionally, we see that the multirate timestepping scheme
does better than the singlerate timestepping scheme for equivalent number of
MPI ranks when $P_c \geq 1024$. This difference is because in the MTS-based
scheme, the interdomain boundary data is exchanged fewer times between the different domains and
each subdomain only needs to re-evaluate extrapolation/interpolation coefficient at
each sub-timestep.  In contrast, STS-based scheme requires data exchange at
each (sub-)timestep.

Based on the results presented here, we conclude that load balance can be
ensured for multirate timestepping-based calculations by choosing the
MPI ranks for each subdomain such that
\begin{eqnarray}
\label{eq:multirateproc}
\frac{P_c}{P_f} \approx \frac{E_c}{\tsr E_f}.
\end{eqnarray}
Though the above relationship might be constrained by the maximum memory
available per MPI rank, as observed in the above test.

\section{Conclusion  }\label{sec:conclusion}
The current work discusses a novel parallel multirate timestepping scheme for the incompressible Navier-Stokes equations in nonconforming overlapping grids. This method scales to an arbitrary number of overlapping grids and is
agnostic of the spatial discretization. The MTS method uses a timestep size based on the local CFL of each subdomain, which unlike the STS-based implementation, avoids unnecessary computation for the subdomain with slower time-scales ($\Omega^c$). The MTS-based framework also requires fewer computational resources for $\Omega^c$ in comparison to STS-based framework because the INSE has to be integrated for fewer timesteps in $\Omega^c$. Using a problem with a known exact solution, we have demonstrated that the MTS method maintains the temporal convergence of the underlying timestepper. We have also demonstrated that the MTS method can accurately model complex turbulent flow using the example of a thermally-buoyant plume. This problem also shows the computational savings associated with the MTS-based method in comparison to an STS-based approach. In future work, we will be extending this method such that the
timestep size and timestep ratios can dynamically change during the calculations based
on the CFL of each subdomain, and accordingly load balance the calculation using \eqref{eq:multirateproc} to further increase the computational savings associated with MTS.

\section{Acknowledgments}
This work was supported by the U.S. Department of Energy, Office of Science, the
Office of Advanced Scientific Computing Research, under Contract
DE-AC02-06CH11357. This research used resources of the Argonne Leadership Computing Facility, which is a DOE Office of Science User Facility. An award of computer time on Blue Waters was provided by the National Center for Supercomputing Applications. Blue Waters is a sustained-petascale HPC and is a joint effort of the University of Illinois at Urbana-Champaign and its National Center for Supercomputing Applications. The Blue Waters sustained-petascale computing project is supported by the National Science Foundation (awards OCI-0725070 and ACI-1238993) and the state of Illinois.

\bibliographystyle{model1-num-names}
\bibliography{lit}

\end{document}